\documentclass[preliminary,copyright,creativecommons]{eptcs}
\providecommand{\event}{MARS 2017} % Name of the event you are submitting to
\usepackage{xspace}
\usepackage[table]{xcolor}
\usepackage{graphicx}
\usepackage{amsfonts,amsmath,amssymb}
\usepackage{enumitem}
\usepackage{wrapfig}
\setitemize{noitemsep,topsep=0pt,parsep=0pt,partopsep=0pt,leftmargin=0.7cm}
\setenumerate{noitemsep,topsep=0pt,parsep=0pt,partopsep=0pt,leftmargin=0.7cm}
\usepackage{awn,awn_keywords}
\usepackage{longtable}

\usepackage{hyperref}

\newcommand{\emptystring}{\varepsilon}

\newcommand{\node}{node\xspace}
\newcommand{\application}{application\xspace}
\newcommand{\component}{component\xspace}

\newcommand{\tcan}{the CAN bus\xspace}
\newcommand{\can}{CAN bus\xspace}
\newcommand{\awn}{AWN\xspace}

\newcommand{\NN}{% natural numbers
    \ensuremath{%
        \mathop{\rm I\mkern-2.5mu N}%
        \nolimits%
    }%
}
\newcommand{\BB}{%
    \ensuremath{%
        \mathop{{\rm I}\!{\rm B}}%
        \nolimits%
    }%
}

\newcommand{\true}{\keyw{true}}
\newcommand{\false}{\keyw{false}}

\newcommand{\myparagraph}[1]{\vspace{1mm plus 1mm}\noindent \textbf{#1.}}

\title{Split, Send, Reassemble:\\
A Formal Specification of a CAN Bus Protocol Stack\thanks{Supported by the Defense
  Advanced Research Projects Agency (DARPA) under agreement number FA8750-12-9-0179.}}
\def\titlerunning{A Formal Specification of a CAN Bus Protocol Stack}
\author{Rob van Glabbeek\qquad\qquad Peter H\"ofner
\institute{Data61, CSIRO, Sydney, Australia}
\institute{School of Computer Science and Engineering\\
University of New South Wales, Sydney, Australia}%
\email{rvg@cs.stanford.edu\quad\qquad Peter.Hoefner@data61.csiro.au}
}
\def\authorrunning{R.J. van Glabbeek \& P. H\"ofner}
\begin{document}
\maketitle

\begin{abstract}
We present a formal model for a fragmentation and a reassembly protocol 
running on top of the standardised CAN bus, which is widely used in  
automotive and aerospace applications. Although the CAN bus
comes with an in-built mechanism for prioritisation, we argue that this is 
not sufficient and provide another protocol to overcome this shortcoming.
\end{abstract}

% !TEX root = ../paper.tex

\section{The CAN Bus Protocol}\label{sec:can}
``A \emph{Controller Area Network (\can)} is a vehicle bus standard designed to allow micro\-controllers and devices to communicate with each other in applications without a host computer.''
\footnote{\url{http://en.wikipedia.org/wiki/CAN_bus}
  (accessed February 23, 2017)}
Robert Bosch GmbH developed it in the 80s and published the latest release in 1991 \cite{CAN}.

The protocol is  message-based and was designed specifically for
automotive applications but is now also used in other areas such as aerospace, maritime and medical equipment.
The {\can} was designed to broadcast many short messages 
to the entire network. The broadcast mechanism provides data consistency in every \node of the system.
Typical information sent are sensor data, such as speed or temperature. 
Due to its simplicity, it is easy to implement; however, its capabilities are rather limited,
in particular w.r.t.\ payload and security.
        
\myparagraph{CAN Bus Limitations}
In the CAN specification, version 2.0, there are two different message
formats to send data in a (typical) CAN network \cite{CAN}.
The only difference between the two formats is that the standard frame format supports a length of 11 bits for the identifier, and the extended frame supports a length of 29 bits.
The payload of both messages is \emph{8 bytes only}.%
\footnote{CAN-FD introduces frames with more than 8 bytes; 
but so far this extension of CAN has not been adopted by the industry.}

CAN is a low-level protocol and offers no (standard) support for any security feature. 
Applications are expected to deploy their own security mechanisms.
Failure to do so can result in various sorts of attacks.
A lot of media attention was generated when cars were hacked and remotely controlled.
The best security mechanism is to ensure that only trustworthy
applications have access to the CAN bus.
An alternative is the use of authentication and encryption, for
instance through HMAC~\cite{rfc2104,rfc6151} or GMAC~\cite{Nist800-38D}.

\myparagraph{The Need for Fragmentation and Reassembly}
As soon as encryption and authentication is implemented, the messages used will be longer than 8 bytes; 
and even without encryption messages are often that long. 
Therefore there is the need for a fragmentation/reassembly protocol. 
In this paper we will present such a protocol  on top of the CAN bus, which remains unchanged.
One reason why we designed our own protocols is that they carry less overhead than 
off-the-shelf solutions.
\advance\textheight 2pt

\myparagraph{The Need for Prioritisation}
The CAN protocol comes with an in-built priority mechanism. 
It uses a bit-wise comparison method of contention resolution, which requires all {\node}s on the {\can} to be synchronised at the point when
transmission begins.

``The CAN specifications use the terms `dominant' bits and `recessive'
bits where dominant is a logical $0$ [\dots] and recessive is a
logical $1$ {[\dots]}. If one \node transmits a dominant bit and
another \node transmits a recessive bit then there is a collision and
the dominant bit `wins'. This means there is no delay to the
higher-priority message, and the \node transmitting the lower priority
message automatically attempts to re-transmit six bit clocks after the
end of the dominant message. This makes CAN very suitable as a real time prioritized communications system.''\footnote{\url{https://en.wikipedia.org/wiki/CAN\_bus} (accessed February 23, 2017)}
A \node that sent a recessive bit and detected a collision ceases transmission and 
will attempt a retransmission of its own message later on.
Since CAN identifiers are unique for each message type and sender,
and constitute the first part of any message, all but one {\node}s will stop while transmitting the CAN identifier. 

\begin{wrapfigure}[18]{r}{0.596\textwidth}
\hfill\includegraphics[width=8.88cm]{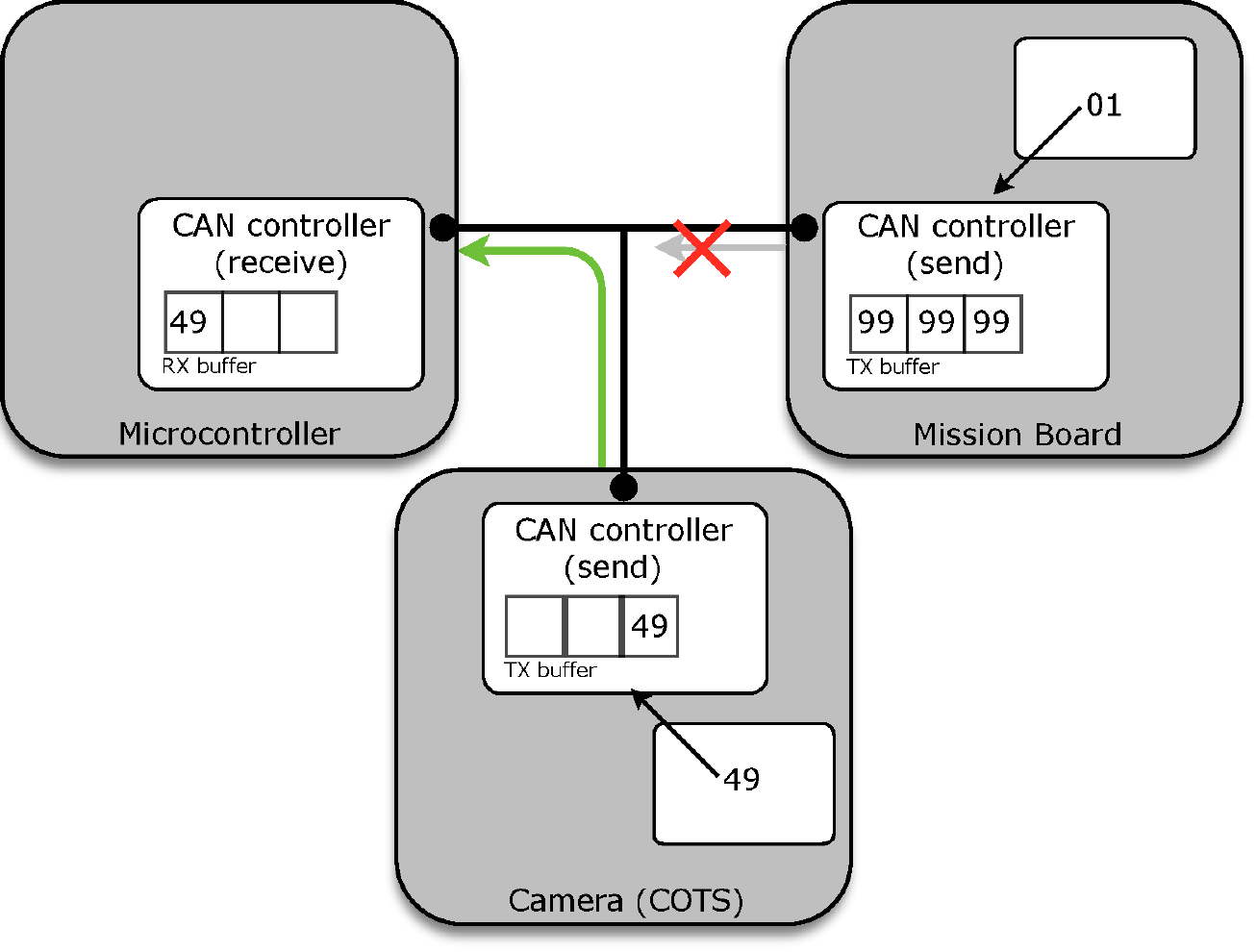} %7
\vspace*{-2mm}
\caption{Blocking Behaviour of High-Priority Messages}
\label{fig:block}
\vspace*{-6mm}
\end{wrapfigure}
While this in-built priority mechanism works if CAN drivers and CAN
controllers\footnote{
A node on a CAN bus is equipped with
a \emph{CAN controller} and a \emph{CAN transceiver}.
The CAN controller has a small number 
(typically 3) of \emph{TX buffers}, where outgoing messages are 
stored before transmission, and a small number of \emph{RX buffers},
which store incoming messages. The CAN transceiver 
manages the actual transmission of messages via the CAN bus---it
  normally sends the message of highest priority stored in a TX buffer first.
The software that  sends messages 
to the controller (which are then stored in the TX buffers), initiates 
the cancellation of messages, and requests messages received, is called a \emph{CAN driver}.}
are considered only, it is not 
always sufficient due to the problem of \emph{priority inversion}. We illustrate this by an example (cf.\ \autoref{fig:block}).   
Assume three {\node}s are attached to the CAN net\-work: a mission board, a microcontroller and a 
camera---this is part of the architecture of our research vehicle (cf.\ \autoref{sec:case}).

Assume that the camera sends a constant stream of messages of
medium priority---say with CAN ID $49$.
This message stream could be interrupted by a message of high
priority, let say $01$, which should be sent from the mission board to the microcontroller. 
Unfortunately, before this really important message is generated, another \application on the mission board generates 3 absolutely unimportant message 
of low priority (here CAN ID $99$). They are transferred to the CAN
controller and stored in the transmission (TX) buffer; 
they are scheduled to be sent via the {\can} as soon as possible. 
Since there is a constant stream of more important messages (the ones
stemming from the camera), these messages are never sent. 
As a consequence the high-priority message cannot be passed on to the
TX buffer, and hence is stuck at the mission board.\footnote{Obviously this example could be avoided by having multiple CAN controllers on the mission board; 
but this cannot be guaranteed and often far more {\application}s
run on a single \node than the number of CAN controllers available.}

This example shows that there is need for another priority mechanism running on each \node separately. 
Such a mechanism should retract one of the low-priority messages from the TX buffer of the mission board and 
replace this message with the high-priority one, which will be sent immediately;
after that the low-priority message will be stored back to the buffer.
There are two possible solutions for such a priority mechanism:
(a) it could be integrated in the CAN driver, or 
(b) it is implemented as yet another protocol that works between the CAN driver and the fragmentation/application layer.
We decided to implement the latter option to have a clear separation of concerns. 

The protocol, called \emph{multiplexer} and formally specified in
\autoref{sec:Mspec}, will be an interface between the CAN driver and 
several instances of the fragmentation protocol. 

\myparagraph{Our Contribution}
We present a protocol stack to be used on top of the \can,
consisting of fragmentation and reassembly protocols, as well as a multiplexer.
It has been formally modelled and partly analysed, implemented, and is 
successfully used in a research vehicle.
The hardware as well as (the implementation of) the CAN bus protocol remains untouched; hence our protocol stack is ready to be deployed on any system featuring a \can.

% !TEX root = ../paper.tex

\section{Our Research Vehicle}\label{sec:case}

\begin{wrapfigure}[7]{r}{0.41\textwidth}
\centering
\vspace*{-5mm}
\includegraphics[width=6cm]{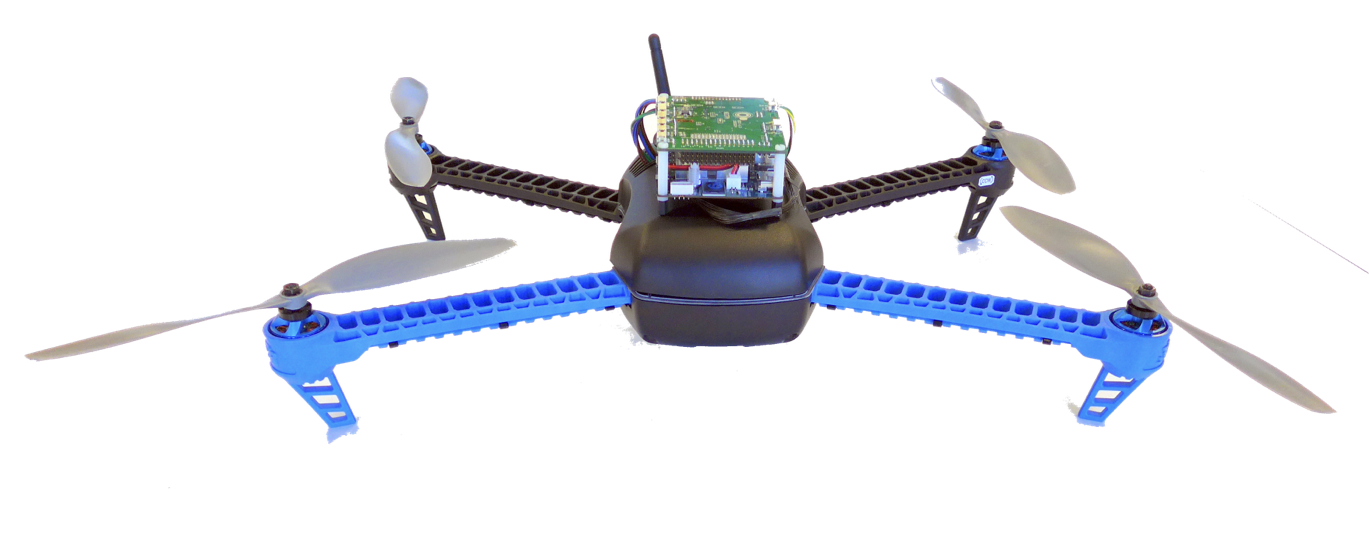}
\vspace*{-6mm}
\caption{Our Research Vehicle
\label{fig:uav}}
\end{wrapfigure}
We use our protocol stack within a research vehicle, namely a small quadcopter. 
This off-the-shelf quadcopter with customised hard- and software was developed as 
part of the SMACCM (Secure Mathematically-Assured Composition of Control Models) project,
a 4.5 year 18 million USD project funded to build highly hack-resilient unmanned aerial vehicles under
\href{http://www.darpa.mil/}{DARPA}'s 
\href{https://opencatalog.darpa.mil/HACMS.html}{HACMS} (High-Assurance Cyber Military Systems)
program.
The team consists of formal verification and synthesis groups in
Rockwell Collins, Data61 (formerly NICTA), Galois Inc, Boeing and the
University of Minnesota.

Our quadcopter is equipped with two boards: a mission board and a control board.
This architecture is artificially made more complex by adding a
trusted gateway and an untrusted COTS (commercial off-the-shelf) component on the bus to introduce some of the
complexities of larger air vehicles.
One of the goals of the project is to show that even if an
  intruder gets hold of the COTS component, or of one of the untrusted applications
  running on the mission board (e.g.\ Linux), this will not invalidate essential
  security properties of the vehicle.

\begin{wrapfigure}[14]{r}{0.705\textwidth}
\vspace*{-6mm}%
\hspace{16pt}\includegraphics[width=11cm]{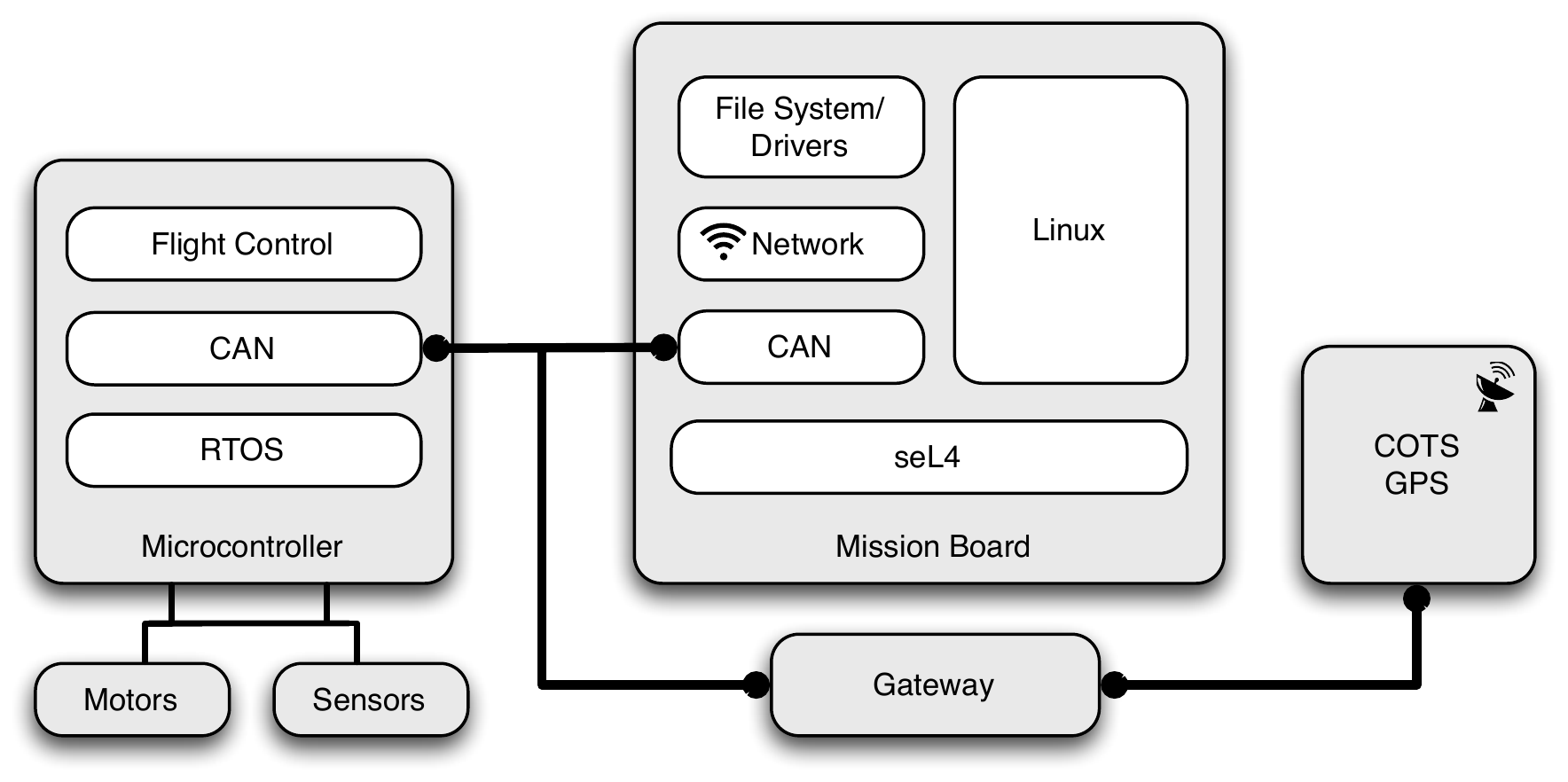}\hspace*{-15pt}%8
\caption{Architecture of the air vehicle with use of \can.}
\label{fig:can}
\end{wrapfigure}
The use of \tcan for communication between the two boards and the COTS
component (via the gateway) is a design decision based on the popularity of CAN in
aviation and automotive applications. It shows that the use of CAN
does not stand in the way of hack-resilience.

\autoref{fig:can} (and in an abstract version also \autoref{fig:block}) shows a sketch of the vehicle's architecture. 
The left hand side shows the flight controller,
which is a pixhawk board with an ARM Cortex M4 CPU; it
has direct connections to
sensors and actuators. The mission board in the central part of \autoref{fig:can} is
more powerful: a TK1-SOM board with an ARM Cortex A15 CPU
with virtualisation extensions running the seL4
microkernel for providing isolation in a mix of trusted and untrusted
applications on top.
The bottom and right-hand-side boxes in \autoref{fig:can} show a gateway
between the trusted part of the internal network on the left and the untrusted
part of the internal network that connects to an unverified component on the right. The purpose of the gateway is to
validate network packets from the right and only let through well-formed
packets to allowed destinations.

The gateway is essential to achieve hack-resilience in the presence of untrusted components.
Giving such components direct access to \tcan might give rise to denial-of-service (DoS) attacks:
a hostile component might supply a continuous stream of high-priority packets, thereby inhibiting
any other traffic. The gateway at least ensures that any message coming from a COTS component has
a CAN identifier that labels it as such. The identifier gives the message a lower priority than safety
critical messages from trusted components. Our multiplexer is designed to ensure that in these
situations safety critical messages cannot be blocked by untrusted components.
It is possible to entrust all security issues to the gateway.
However, a simpler gateway need not investigate and restrict information flowing towards the COTS
component; in this case safety critical messages that need to be kept private could be encrypted.

For the protocols presented in this paper and used in our research vehicle, we assume a functionally correct \can. The
proposed architecture and kernel-provided isolation on the mission board
minimises the potential attack surface compared to standard systems. 
Messages that need splitting arrive from a ground station---which plans and manages missions---to the mission board and 
are forwarded to the microcontroller; the messages itself are often longer than 8 bytes, and when encrypted exceed definitely the payload of a standard CAN message. 

% !TEX root = ../paper.tex

\section{Assumptions and Requirements}\label{sec:assum}
During the design of the fragmentation protocol we had to make some assumptions. 
All assumptions are realistic and can be assumed for our case study without loss of generality.

\myparagraph{Assumptions on the CAN Bus}
\newcounter{mycount}

\begin{enumerate}
\item For the verification of two central correctness properties
  of the protocol---any message received has been sent, i.e.\ split
  messages are not reassembled in the wrong way, and any message sent
  is received---we assume a perfect channel and assume that
  \emph{every message sent via \tcan will be received} by all
  {\node}s that are connected to the \can.
However, for our more basic correctness properties, such as the
  absence of deadlocks in protocol components or in the entire
  protocol stack, and the unreachability of error states, we do not
  make such an assumption.

\begin{enumerate}

\item
The CAN protocol specifies an automatic retransmission of faulty messages after transmitting an error frame. 
Error frames may be sent by transmitting or receiving nodes. 
This happens on a lower protocol layer than the one modelled here.

\item
  In case resending of an entire message is needed (e.g.\ if fragments
  are lost), we leave this task to the application layer/user. 
The reason for this decision is that most information sent via the
\can is time-sensitive; so if the information is not sent immediately, 
it will be outdated. Examples are GPS-coordinates or telemetry data.
\end{enumerate}
\item We allow for the possibility that a CAN message is sent and received
  twice---possibly because one of the receivers used the error frame
  to ask for a resend. Our protocol is required to deal with such a
  repeated fragment.

\item Messages sent over \tcan are \emph{not reordered}.
This is a realistic assumption.
The CAN protocol sends one packet after the other; fragmented messages stemming from different sources may be interleaved, but reordering of messages sent by a single 
sender does not occur.
Even if a packet is lost, the resending happens before the next frame is processed and sent.

A CAN controller allows overtaking of low priority messages by high
  priority messages by first offering the highest priority message stored in
  its TX buffers for transmission. The above assumption therefore only
  rules out reordering after messages have been transmitted on the \can.

\setcounter{mycount}{\value{enumi}}
\end{enumerate}

\myparagraph{Requirements on the Input Data}
\begin{enumerate}
\setcounter{enumi}{\value{mycount}}

\item
Every message type, as determined by a CAN identifier (ID), has a
\emph{unique sender}.\footnote{Even if two {\application}s
  would send messages of the same message type, one could use two different IDs.}
This requirement is reasonable and reflects the CAN protocol in the way it is used in the automotive industry.
As a consequence it is impossible to have two senders sending messages with the same ID at the same time, 
and hence collisions are avoided.\footnote{The in-build priority
messages will take care of messages with different IDs sent at the same time (see \autoref{sec:can}).}
By using wrong identifiers this requirement can be violated, yielding
collisions and message loss. 
Thus it should be ensured that this requirement is satisfied.
\item Every message we are going to split has a \emph{fixed length}, determined by its message type. As a consequence we know in advance into how many fragments a message needs to be split. 
In case the message lengths can vary we assume that there is an upper bound and that shorter messages are extended by `dummy bits'.

\setcounter{mycount}{\value{enumi}}
\end{enumerate}

\myparagraph{Requirements for our Fragmentation Protocol}
\begin{enumerate}
\setcounter{enumi}{\value{mycount}}

\item The \emph{distribution of the message IDs} is fixed at compile time. We do not restrict the choice, as long
as the assumption `unique sender for every message ID' is
maintained,  and adjacent IDs  are allocated to adjacent fragments of messages of the same type.

\item We assume that the application layer, after submission of a
  message to an instance of the splitting protocol for transition over the \can,
  will wait for an acknowledgement (positive or negative) before
  submitting a new message to that instance of the splitting protocol. The application layer may at any time
  submit a cancellation request for the last message submitted; this
  will speed up the (now likely negative) acknowledgement.

\item Our protocol has to support \emph{legacy {\node}s}. 
Independent of the software {\component}s we are adding (fragmentation protocol, authentication, encryption), the original CAN protocol 
should still be available and it must be possible to send and receive ordinary CAN messages.
\vspace{1.5ex}
\end{enumerate}

\noindent
Our fragmentation protocol uses a new CAN identifier for each fragment of each message type.
One might wonder if this does not lead to a shortage of CAN IDs. Within our research vehicle this
problem did not occur, and experts from the automotive side have ensured us that in typical
applications there is an abundance of unused CAN IDs, in
particular when using the extended frame format with 29-bit
identifiers that allow over half a billion identifiers.

% !TEX root = ../paper.tex
\vspace{-1mm}
\section{Related Work}
\vspace{-1mm}
Several fragmentation protocols to be deployed on top of \tcan have been developed in the past.

The \emph{ISO-TP}~\cite{isotp} or \emph{ISO 15765-2} protocol is an international standard for sending data packets over a \can that exceed the 8 byte maximum payload.
ISO-TP splits longer messages into multiple frames, adding metadata
that allows the interpretation of individual frames and reassembly
into a complete message packet by the recipient. The protocol can handle message of up to 4095 bytes.
The first fragment can only carry up to 7 bytes when using \emph{normal addressing}, instead of the standard 8 bytes. 
Hence every message longer than 7 bytes should be split; as a consequence special care has to be taken so that this protocol can handle legacy messages.
\pagebreak

Shin~\cite{Shin14,ShinPatent} follows the spirit of ISO-TP and describes a protocol similar to ours.
Due to his design the payload of each CAN message is split up into an 8-bit message identifier, a 7 bit sequence number, which points to the next fragment, and 
the actual payload, which has a maximum capacity of 6 bytes---2 bytes less than in original CAN frames and in our approach. 
As a consequence we have far less overhead than Shin's approach. 
Moreover, his reassembling routine does not take packet loss into account.

The \emph{TP 2.0} protocol, sometimes also called \emph{VW TP 2.0}, (see e.g.\ \cite{SD15}) introduces a connection-oriented approach. It first sends a couple of messages to establish a ``channel'' between 
the sender and the recipients, then exchanges and sets up channel parameters such as the number of frames (fragments) to be sent, and finally transmits the message (by encoding the channel number into the CAN ID).
Although the used CAN frames can use the full pay load of 8 bytes, the overhead lies in the setup-phase.

The \emph{Service Data Object} (SDO) protocol,\footnote{\url{http://www.can-cia.org/can-knowledge/canopen/sdo-protocol/}} which is part of CANopen\footnote{\url{http://www.can-cia.org/can-knowledge/canopen/canopen/}}
 (see e.g.\ \cite{E01})---a communication protocol for embedded systems used in automation---also  implements segmentation and desegmentation of longer messages.
SDO is used for setting and for reading values from the object dictionary of a remote device. 
Because the values can be larger than the 8 bytes limit of a CAN frame, the SDO protocol implements also a fragmentation protocol.
However, the fragmentation itself is a subprotocol and using SDO or even the entire CANopen infrastructure is again too much of an overhead.

All of these protocols bear a more or less heavy overhead w.r.t. the number of messages. Moreover,
none of these solutions appear compatible with the priority mechanism offered by our multiplexer.
In fact, adopting any of them stands in the way of inserting a multiplexer between the
fragmentation protocol and the CAN driver, thus taking care of priority inversion, as we do here.
Therefore, we designed our own simple protocol that exactly meets our needs; by 
using the vast amount of CAN identifiers available, our protocol does not have any overhead w.r.t.\ the number of messages sent.

% !TEX root = ../paper.tex

\section{Protocol Stack: Overall Structure and Informal Description}\label{sec:overview}
Before presenting our formal specifications, including formally defined data structures and 
unambiguous process-algebraic specifications, we describe the overall
structure of our protocol stack, including 
all messages sent between {\component}s. 
By abstracting from formal details we
discuss the `big picture'.

\begin{figure}
\centering
\includegraphics[width=10.7cm]{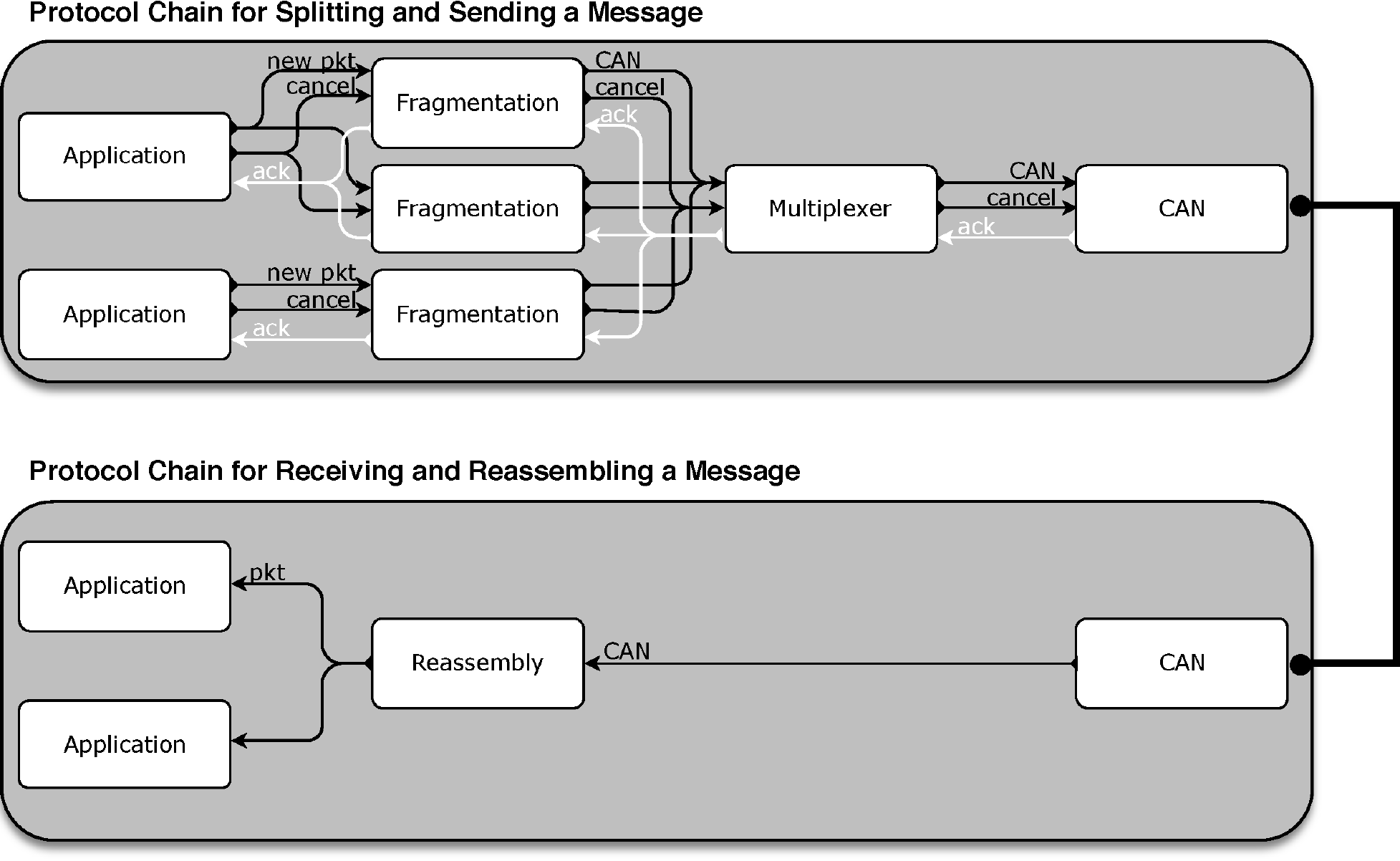}
\vspace*{-2mm}
\caption{Message passing between the different {\component}s}
\label{fig:overview}
\vspace{-4mm}
\end{figure}

Figure~\ref{fig:overview} gives a schematic representation of our protocol stack. 
Every hardware component (\node) contains exactly one instance of each of the two protocol chains.
We distinguish 4 different layers. 
\begin{itemize}
\item The application layer: applications are {\component}s that send messages to and receive messages from the {\can} (via the other 3 layers).
\item The CAN layer: this layer combines the CAN controller and the CAN driver. We will model a simple instance of this layer.
\end{itemize}
\noindent The remaining two layers connect the application with the CAN layer---they split and reassemble messages if necessary, and handle 
messages with high priority first.
\begin{itemize}
\item The fragmentation/reassembly layer: 
the fragmentation protocol receives a message from an application, and, depending on the type of the message,
simply forwards the message if it is a short or legacy, or otherwise  splits the message into 
fragments of 8 bytes each. These fragments have the form of standard CAN messages and can (now) be sent via the {\can}. 
The reassembly protocol receives such fragments that were sent via the {\can} and forwarded by the CAN driver. It stores the fragments until a full message can be reassembled and transmitted to the application. 
\item The multiplexer accepts messages from different instances of the fragmentation protocol (all running on the same hardware) and 
stores them in a priority queue. It always sends the message that needs
to be transmitted next to the CAN
driver, if necessary after cancelling a lower-priority message sitting in the
  TX buffer of the CAN controller. 
This prevents the blocking example of \autoref{sec:can}. 
\end{itemize}
All these layers exchange information through message passing. In the
remainder of the section we will
discuss the different types of messages that occur in our protocol chain. 
{%redefine message names (for simple overview)
\renewcommand{\cancelID}{\keyw{cancel}}
\renewcommand{\canceliID}{\keyw{cancel}}
\renewcommand{\ackID}{\keyw{ack}}
\renewcommand{\canmID}{\keyw{can}}

\myparagraph{Protocol Chain for Splitting and Sending a Message}
Any application is allowed to submit \emph{new messages}
$\newpktID$. Every such message
contains a message type and a payload. 
Depending on the type the message is sent to a particular instance of the fragmentation protocol. 
For this we assume that
  \emph{for each message type there is exactly one instance of the fragmentation protocol};
an instance, however, is allowed to handle multiple message types,
provided that they all stem from the same application.
An application usually sends one new message at a time; in case a
second packet is injected before the previous one has been fully handled, the protocol yields an
error and deadlocks. To avoid this scenario the fragmentation protocol returns an acknowledgement
message $\ackID$, informing the application that the handling of the message is finished---for each
$\newpktID$ received exactly one $\ackID$ is sent.
The acknowledgement can be positive or negative, depending on whether the message was
successfully sent via the {\can} or not.
The application has also the possibility to inject a
$\cancelID$-message to the fragmentation protocol, which will then stop handling the last message injected
by the application, unless it is already finished with it.

The fragmentation protocol receives new messages from the application and splits them; the resulting
CAN messages are passed on to the multiplexer.  As in case of an application, the fragmentation
protocol awaits an acknowledgement $\ackID$ (positive or negative) before sending the next CAN
message.  In case the fragmentation protocol receives a cancellation request, it stops splitting a
message, and in case the multiplexer has not yet acknowledged the last fragment submitted to it also
informs the multiplexer about the cancellation request by sending it a $\canceliID$-message.  The
multiplexer returns exactly one acknowledgement for each fragment received; it does not accept a
second fragment from the same fragmentation instance before this $\ackID$ has been sent. It accepts
a cancellation request at any time.
%, and erases the corresponding fragment from its priority queue if it has not already been handled.

The multiplexer schedules all messages received in an appropriate order (high priority messages first); the messages involved in this are again $\canmID$, $\canceliID$, 
and $\ackID$. The CAN-messages that were sent by the multiplexer are handled by the CAN layer and end up in the TX buffer of the CAN controller, from where they are transmitted over the {\can}.
\pagebreak

\myparagraph{Protocol Chain for Receiving and Reassembling a Message}
After the messages were transmitted via the {\can}, they are stored in the RX buffer of the CAN controller. 
The CAN layer stores and handles all incoming messages of type $\canmID$ and forwards them to the reassembly protocol,
which  stores these messages---there is no pendant to the multiplexer in the receiving chain. As
soon as a packet is fully received and reassembled, it is delivered to the appropriate application,
using a message of type~$\fndatamsg$.
}%end redefinition

A formal specification of the multiplexer is presented in \autoref{sec:Mspec}. Formal specifications of all
the above-mentioned protocols are
given in \autoref{app:spec}. This includes the
processes themselves, specified in the process algebra \awn (see below), as well as a detailed definition of the data structure involved. 

% !TEX root = ../paper.tex

\section[AWN: A Specification Language for Protocols]{AWN: A Specification Language for Protocols\footnotemark}\label{sec:awn}
\footnotetext{Parts of this section are published in~\cite{GHPT16}.}
Ideally, any specification is free of ambiguities and
contradictions. Using English prose only---as is still state of the
art in protocol specification---this is nearly impossible to achieve.
The use of {\em any} formal specification language helps to avoid ambiguities and to precisely describe the intended behaviour.
The choice of a specification language is often secondary, although it has high impact on the analysis. 

For this paper we choose the modelling language \awn~\cite{ESOP12},
which provides the right level of abstraction to model key protocol features, while abstracting from implementation-related details. As its semantics is completely unambiguous, specifying a protocol in such a framework enforces total precision and the removal of any ambiguity.
{\awn} is tailored for modelling and verifying routing and communication protocols and therefore offers primitives such as {\bf unicast} and {\bf multicast}/{\bf groupcast}; 
it defines the protocol in a pseudo-code  that is easily readable---the language itself is implementation independent; and
it offers some degree of proof automation and proof
verification~\cite{BGH16,BGH14b}, using Isabelle/HOL~\cite{NipkowPaulsonWenzel02}.

\awn is a variant of standard process algebras, extended with a local
broadcast mechanism and a novel \emph{conditional unicast}
operator---allowing error handling in response to failed
communications while abstracting from link layer implementations of
the communication handling---and incorporating data structures with
assignments; its operational semantics is defined in \cite{ESOP12}.

We use an underlying data structure (described in detail in
\autoref{app:spec}) with several types, variables
ranging over these types, operators and predicates. First
order predicate logic yields terms (or \emph{data expressions}) and
formulas to denote data values and statements about them.
Our data structure
always contains the types \tDATA, \tMSG, {\tIP} and $\pow(\tIP)$ 
of
\emph{application layer data}, \emph{messages}, \emph{identifiers}
and \emph{sets of identifiers}.
The messages comprise \emph{data packets}, containing application layer
data, and \emph{control messages}.

In \awn a network is modelled as a parallel composition of
{\component}s. Here any party in the network that can be
addressed as the recipient of a message is a \component.
On each \component several processes may be running in parallel.
Components communicate with their direct neighbours, which in our
current application are all other {\component}s in the network.

The \emph{Process expressions} are given
in \autoref{tb:procexpr}.
{\footnotesize
 \begin{table}[t]
 \centering
{\footnotesize
  \setlength{\tabcolsep}{2.6pt}
 \begin{tabular}{|l|l|}
\hline
\rule[6.5pt]{0pt}{1pt}%
$X(\dexp{exp}_1,\ldots,\dexp{exp}_n)$& process name with arguments\\
$P+Q$ & choice between processes\ $P$ and $Q$\\
$\cond{\varphi}P$&conditional process: proceed as $P$, but only if
  $\varphi$ evaluates to \true\\
$\assignment{\keyw{var}:=\dexp{exp}}P$&assignment followed by process $P$\\
$\broadcast{\dexp{ms}}.P $&broadcast message \dexp{ms} followed by $P$\\
$\groupcast{\dexp{dests}}{\dexp{ms}}.P$&
  multicast \dexp{ms} to all destinations \dexp{dests}\\
$\unicast{\dexp{dest}}{\dexp{ms}}.P \prio Q$& unicast $\dexp{ms}$ to $\dexp{dest}$; if successful proceed with $P$; otherwise with $Q$\\
$\deliver{\dexp{data}}.P$&deliver data to application layer\\
$\receive{\msg}.P$&receive a message and store its contents in
    the variable \msg\\
\hline
\rule[6.5pt]{0pt}{3pt}%
$(\xi,P)$   & process $P$ with initial valuation of its variables\\
$V \parl W$ & parallel valuated processes on the same \component\\
\hline
\rule[6.5pt]{0pt}{3pt}%
$\dval{id}\mathop{:}V$ & addressed {\component}\\
$C\|D$ 	&parallel composition of addressed {\component}s\\
\hline
\end{tabular}}
\caption{Process Expressions}
\label{tb:procexpr}
\end{table}
}%end small
They should be understandable without further explanation; 
we add a short description in \autoref{app:AWN}.

% !TEX root = ../paper.tex

\section{A Formal Specification of the Multiplexer}\label{sec:Mspec}

In this section we present two out of four {\awn}-processes that entirely specify our multiplexer. 

\myparagraph{The Main Loop}
The basic process \MULT\ (\Pro{multiplexer_intext}) receives
messages from the fragmentation protocol or the CAN driver. 
Since the multiplexer is not always ready to receive messages, we equip the process with an in-queue (see \autoref{app:spec});
so technically the multiplexer receives a message from this queue.
\MULT\ maintains two data variables {\pqueue} and {\txmirror}.
The former implements a priority queue which contains all CAN messages to be sent via 
the {\can};
the later is a local storage which keeps track of the CAN IDs
currently sent by or stored in the TX buffers of the CAN controller. 

\newcommand{\multilabel}[1]{\label{multi:#1}}
\setcounter{algorithm}{8}
  \algsetup{linenodelimiter=.,linenosize=\tiny}
  \begin{algorithm}[ht]
    {\footnotesize
      \caption[]{Multiplexer---Main Loop$^{\ref{fn:numbering}}$}
      \label{pro:multiplexer_intext}
      \begin{algorithmic}[1]
        % !TEX root = ../paper.tex
\DEFPROCESS{\MULT}{\pqueue\comma\txmirror}													\multilabel{line0} 
	\receiveL{\msg}\ .                         																			\multilabel{line1}                                                   
	\PAR                         																								\multilabel{line2}      
	\IF[new fragment]{$\msg = \canm{\cid}{\data}$}     													\multilabel{line3}     
		\treqL{\msg}{\pqueue}{\txmirror}																			\multilabel{line4}      
	\ELSIF[cancellation message received]{$\msg = \canceli{\cid}$} 							\multilabel{line5}      
		\ccancL{\cid}{\pqueue}{\txmirror}																		\multilabel{line6}      
	\ELSIF[message from CAN controller]{$\msg = \amsg{\txid}{\ackM{\ackb}}$}			\multilabel{line7}      
		\ackcL{\ackb}{\txid}{\pqueue}{\txmirror}																\multilabel{line8}      
	\ENDIFii
	\ENDPAR																												\multilabel{line9}      

	\end{algorithmic}
    }%end{footnotesize}
  \end{algorithm}
\renewcommand{\multilabel}[1]{\label{appmulti:#1}}

First, a message has to be received (Line~\ref{multi:line1}). 
After that, the process \MULT\ checks the type of the message and calls a process that can handle this message: 
in case a CAN message is received from the fragmentation protocol, the process
\TREQ\ is called (Line~\ref{multi:line4});  
in case of an incoming cancellation request the process \CCANC\ is executed (Line~\ref{multi:line6});
and in case a message from the CAN driver is read, the process \ACKC\ is called (Line~\ref{multi:line8}).
%, carrying the Boolean flag {\ackb} and the name of the TX buffer ($\txid$) received from the CAN driver.
In case a message of any other type is received, the process
\MULT\ deadlocks; it is a proof obligation to check that this will not occur.

\myparagraph{New CAN Message}
In case a new CAN message is sent from an instance of the fragmentation protocol, 
the process {\TREQ} stores the CAN message and determines whether the newly received message 
is important enough to be forwarded directly to the CAN driver. The formal specification is 
shown in \Pro{newcan}.

\newcommand{\treqlabel}[1]{\label{treq:#1}}
  \algsetup{linenodelimiter=.,linenosize=\tiny}
  \begin{algorithm}[ht]
    {\footnotesize
      \caption[]{New CAN Message Received\footnotemark}
      \label{pro:newcan_intext}
      \begin{algorithmic}[1]
        % !TEX root = ../paper.tex

\DEFPROCESS{\TREQ}{\msg\comma\pqueue\comma\txmirror}																			\treqlabel{line0}
	\IF[distill $\cid$ out of $\msg$]{$\msg = \canm{\cid}{\data}$}
	\UPD{\pqueueupd{\cid} := \msg} 	\COM{store message in priority queue}															\treqlabel{line1}
	\PAR
	\IF[message should be scheduled]{$\cid\in\nbest{\pqueue}$}																				\treqlabel{line2}
		\PAR																																								\treqlabel{line3}
		\IF[TX buffer $\txid$ is free]{$\txcid{\txid}{\txmirror} = \cidundef$}																	\treqlabel{line4}	
			\UPD{\mboxupd{\txid}:=(\cid,\false)}  																											\treqlabel{line5}
			\unicastL{\C}{\amsg{\txid}{\msg}}\ .\  \COM{pass message to CAN	driver, to put in free slot}						\treqlabel{line6}
			\multiL{\pqueue}{\txmirror}																															\treqlabel{line7}
		\ELSIF[cancel message with lowest priority]{$\forall \txid\in\tTX: \txcid{\txid}{\txmirror} \not= \cidundef$}			\treqlabel{line8}
			\PAR																																							\treqlabel{line9}
				\UPD{\wtx := \worstmbid{\txmirror}}	\COM{identify TX buffer containing lowest CAN ID}						\treqlabel{line10}
				\PAR
				\IF[TX buffer $\wtx$ is still active]{$\txab{\wtx}{\txmirror} = \false$}															\treqlabel{line11}
                    \UPD{\mboxupd{\wtx}:=(\txcid{\wtx}{\txmirror},\true)}  \COM{set the abort-flag of buffer $\wtx$}            \treqlabel{line12}
					\unicastL{\C}{\amsg{\wtx}{\cancel}}\ .\		\COM{cancel contents of buffer $\wtx$}								\treqlabel{line13}
					\multiL{\pqueue}{\txmirror}																													\treqlabel{line14}
				\ELSIF[TX was already asked to clean up]{$\txab{\wtx}{\txmirror} = \true$}												\treqlabel{line15}
					\multiL{\pqueue}{\txmirror}																													\treqlabel{line16}
				\ENDIFii
				\ENDPAR						
			\ENDPAR																																					\treqlabel{line17}
		\ENDIFii
		\ENDPAR																																						\treqlabel{line18}
	\ELSIF[message not important enough to be scheduled right now]{$\cid\not\in\nbest{\pqueue}$}								\treqlabel{line19}
		\multiL{\pqueue}{\txmirror}																																\treqlabel{line20}
    \ENDIFii
    		\ENDPAR
    \ENDIFii

	\end{algorithmic}
    }%end{footnotesize}
  \end{algorithm}
\renewcommand{\treqlabel}[1]{\label{apptreq:#1}}

\footnotetext{\label{fn:numbering}The numbering of the processes is according to \autoref{app:spec}.}

\setcounter{algorithm}{0}

The received CAN message is first stored in the queue {\pqueue} (Line~\ref{treq:line1}), which contains all messages to be sent via the 
{\can}. The protocol just stores the newly received message, it does not check for emptiness 
of $\pqueueupd{\cid}$. Therefore, to guarantee that no message is lost 
the property $\pqueueupd{\cid}=\msgundef$ needs to hold before Line~\ref{treq:line1} is executed;
it needs to be proven.  
The protocol then determines whether the message should directly be forwarded to the CAN driver---this is the case 
if the CAN ID is among the $n$ messages with lowest CAN IDs currently stored in {\pqueue} (Line~\ref{treq:line2}). Here $n$ equals 
the number $\noTX$ of TX buffers available in the CAN controller. 
Lines~\ref{treq:line3}--\ref{treq:line18} present all actions to be performed in case the message 
is forwarded to the CAN driver. 

In case there exists an empty TX buffer $\txid$, which is currently not used,
the message should be sent to this TX buffer, and there is no need to erase a used TX buffer. 
The empty buffer $\txid$ is chosen in Line~\ref{treq:line4}.\footnote{Since $\txid$ is a free variable, 
it will be instantiated with a value that validates
$\txcid{\txid}{\txmirror} = \cidundef$;
so the condition in the guard is satisfied iff $\exists \txid\in\tTX: \txcid{\txid}{\txmirror} = \cidundef$.}
The CAN message is then forwarded to the connected CAN driver {\C} in Line~\ref{treq:line6}. Since the CAN driver needs also the 
name of the TX buffer to be used, the value $\txid$ is sent next to the CAN message $\msg$.
The multiplexer also updates the local variable $\txmirror$ (Line~\ref{treq:line5}), which keeps track 
of those CAN identifiers that are currently sent by or stored in the TX buffers. By this, the newly received message 
has been handled and the process can return to the main routine (Line~\ref{treq:line7}).

In case all available TX buffers are used (Line~\ref{treq:line8}), the least important message---the CAN message with the largest
CAN ID---needs to be removed from the TX buffer and rescheduled later. 
This avoids the blocking example presented earlier.
In Line~\ref{treq:line10} the process {\TREQ} determines the name of the TX buffer that contains the `worst' message currently 
handled for sending. The CAN message that should be stored in this
particular TX buffer cannot be put there immediately; a cancellation request needs to be sent first, 
and an acknowledgement needs to be received that informs the multiplexer about a free TX buffer.
The routine checks whether a cancellation request was sent earlier, using the 
function $\fntxab$. If this is the case, it returns straight to the process \MULT; otherwise 
a cancellation message is sent to the CAN driver \C, identifying the TX buffer that needs cancellation (Line~\ref{treq:line13}).

If the newly received CAN message is not important enough to be forwarded to the CAN driver immediately 
(Line~\ref{treq:line19}), the process {\TREQ} just returns to the main process (Line~\ref{treq:line20}), where it
awaits a new message. The stored message will be handled later when a TX buffer becomes available.

% !TEX root = ../paper.tex
\newcommand{\ERR}{\keyw{ERR}}
\newcommand{\errP}{\ERR\ensuremath{()}}

\section{Properties and Formal Analysis}
After defining the splitting and reassembly protocol in a formal and unambiguous manner (including the multiplexer), 
we can now focus on verification tasks. A detailed analysis and verification is out of the scope of this paper---which 
concentrates on the necessity of the protocols and their formal specification. In this section we list a series of desired properties---a first analysis using model checking techniques indicates that these properties are satisfied.%
\footnote{In fact we did find an error in the multiplexer that has been eliminated in the current version.}

\myparagraph{Unreachability of {\err} State}
In our specification we use a special state  {\err} that is reached by a
component process whenever it receives an input that is unexpected, and for which no
proper response in envisioned. As {\err} is not an {\awn} primitive, we\vspace{-2pt}
proposed to simply implement it in terms of {\awn} primitives as a deadlock:
$\keyw{ERROR}() \stackrel{{\it def}}{=} \cond{\false}\,\keyw{ERROR}()$.

Our first requirement on the correctness of the overall protocol is
that none of its components will ever reach the {\err} state.
Since the application layer is not part of our specification, we cannot 
show that it behaves properly. Instead, we have to formulate a
requirement on the communications between the application layer and
our protocol; we only require the unreachability of {\err} under that condition.

\myparagraph{The Protocol is Deadlock Free}
Another requirement is that our protocol is deadlock free, in the
sense that each reachable state has an outgoing transition.
As termination of our protocol is not envisioned, a deadlock is a
clear case of undesirable behaviour.
This requirement does not rule out any state where no further activity
occurs due to lack of input from the application layer, for the
possibility of such input is modelled as an outgoing transition.

\myparagraph{Each Component of the Protocol is Deadlock Free}
A related requirement is that each component in our specification is
deadlock free. The components are all instances of the fragmentation protocol, the
multiplexer, the CAN receiver, the CAN sender, and the reassembly
protocol. Optionally, the input queue of the multiplexer can be
regarded a separate component, too.

This requirement is neither weaker nor stronger than the above
requirement that the entire protocol is deadlock free, for it
could be that a deadlock of the entire protocol occurs when two
components fail to properly communicate, even though each of them
could make progress if only the other one cooperated.

\myparagraph{Any Message Received Has Been Sent}
Here a message counts as `sent' when it is submitted by the
application to the fragmentation protocol;
it counts as `received' when it is passed on by the reassembling
protocol at each node listed as a destination of the message to the
corresponding application layer.

This requirement is definitely  violated in case our reassembly protocol reassembles
messages the wrong way. Such a situation can occur in case of message
loss on the CAN bus. 
%For suppose that the application layer sends two
%messages A and B of the same message type in succession. The
%fragmentation protocol splits each message in three parts, and thus
%forwards A$_1$, A$_2$, A$_3$, B$_1$, B$_2$, B$_3$ to the multiplexer,
%which passes them on to the CAN bus. Now assume that A$_3$, B$_1$ and
%B$_2$ disappear without a trace. Then the CAN receiver will forward
%A$_1$, A$_2$ and B$_3$ to the reassembly protocol, which then sorts them by message
%type (thus in the same bucket) and pastes them together as one message
%of the appropriate length, which is forwarded to the application layer
%at the other end.

For this reason, we can only hope to establish the property under the
condition that no message loss occurs. In our model, this is quite
simple, as the possibility of message loss is not modelled. This
follows the advice of experts on the CAN bus, saying that any message
that is actually sent from a TX buffer at the transmitting end is
always picked up by a RX buffer at the receiving end.

%When message loss does occur, the possibility of violation of the
%requirement that no message shall be received by the application layer
%that has not been sent by an application, is mitigated by the use of
%encryption. Namely if an encrypted message is pasted together the wrong
%way, the probability that it can be decrypted into a valid message is
%basically zero. If we do not count a message that does not decrypt as one
%that is properly received, the requirement should be achieved with very high
%probability. 

\myparagraph{Any Message Sent Is Received}\label{received}
This may be regarded as the central requirement of the protocol.
In fact, some of the requirements above are in some sense entailed by
this requirement, as the presence of deadlocks will surely
manifest itself as failure to handle and receive further messages.

Obviously, this requirement cannot be guaranteed if there is message
loss on the CAN bus. Thus, as before, we assume
that no loss on the CAN bus occurs. 
%The requirement itself now says that there is no loss anywhere else
%either. Note that in case message loss on the bus does occur,
%encryption will not offer the solution it provides for Requirement~\ref{sent}.
Since the CAN bus handles higher
priority messages in preference to lower priority messages, a given
message that is submitted to the protocol by the application layer at
a node may never be selected by the CAN bus if a steady stream of
higher-priority messages is send by another node; so this property does not hold. 
Consequently, further the assumption that there is no steady stream of higher priority messages is required.

The property is also violated if the input queue of
the multiplexer is grossly unfair, in the sense that it has no time to
accept a message from one of the fragmentation processes because it is
continuously busy accepting incoming messages from other fragmentation
processes on the same node. 
Since the input queue of the multiplexer is an order of magnitude faster
than the CAN bus, there should be no situation where there is contention
in getting into that queue.

One of the main dangers faced by the CAN bus and its surrounding
protocols is a Denial of Service (DoS) attack. 
In our application (see \autoref{sec:case}), we protect the bus against DoS attacks by giving
only trusted software access to a (secure) CAN bus.
Unsecure components may also send messages to the bus, but those
message first pass through a trusted gateway, which performs rate limiting.

The proof of this requirement will undoubtedly be the
hardest part of the verification effort. It probably requires various
intermediate results, such as
`\emph{every fragment sent by a fragmentation process is received by the
  reassembly process on each of its destination nodes}'.

\myparagraph{Buffers Have a Maximal Length}
Our formal specification employs message buffers in two places.
One is (in) the CAN receiver; the other is the input queue of the multiplexer. Both buffers are modelled as FIFO queues.
Following the specification both have unbounded capacity.
In reality, buffers have a bounded capacity, and overflows will occur
when trying to exceed it.

For the CAN receiver an upper bound cannot be given
without a timing analysis comparing the capacity of the CAN bus 
and the `working speed' of the reassembly protocol.

The input queue of the multiplexer, on the other hand, 
has a maximal length. 
To calculate this length, we add the maximal
number of messages it could receive from any fragmentation process,
times the number of fragmentation processes running on the node,
plus the maximum number of message received from the associated CAN controller.
For each type of message we calculate an upper bound.

\myparagraph{The Application Layer Can Always Succeed in Submitting a New Message}
A requirement mentioned above guarantees that, under certain conditions,
messages submitted will eventually reach their destinations.
 As a liveness property for the application layer, this is only convincing if in
addition the application layer can always succeed in submitting to the
fragmentation protocol any message its want to transmit.

% !TEX root = ../paper.tex

\section{Conclusion and Future Work}
In this paper we have presented a formal and unambiguous specification of a fragmentation and reassembly protocol, 
as well as a multiplexer. These protocols are running on top of a standard CAN bus and hence can directly be 
applied in  many areas such as the automotive space. We have argued why both protocols are needed in 
real applications; and have shown applicability by running our protocols on a research quadcopter. 

Last but not least we have listed a couple of important properties our protocol stack should satisfy. 
It is our belief that the presented properties do hold for our formally specified protocols, 
at least under some assumptions (as we have pointed out). Future work could provide formal proofs for these properties.

\bibliographystyle{eptcs}
\bibliography{refs}

\newpage
\appendix
% !TEX root = ../paper.tex
\renewcommand{\node}{\textcolor{blue}{component}}

\section{Informal Description of the Process Expressions of {\awn}}\label{app:AWN}
In this appendix we describe the \emph{Process expressions}, given
in \autoref{tb:procexpr}.
 
A process name $X$ comes with a \emph{defining equation}\vspace{-1ex}
\[X(\keyw{var}_1,\ldots,\keyw{var}_n) \stackrel{{\it def}}{=} P\ ,\]
where $P$ is a process expression, and the $\keyw{var}_i$ are data
variables maintained by process $X$. A named process is like a
\emph{procedure}; when it is called, data expressions $\dexp{exp}_i$
of the appropriate type are filled in for the variables $\keyw{var}_i$.
Furthermore, $\varphi$ is a condition,
$\keyw{var}\mathop{:=}\dexp{exp}$ an assignment of a data expression
\dexp{exp} to a variable \keyw{var} of the same type, \dexp{dest},
\dexp{dests}, \dexp{data} and \dexp{ms} data expressions of types
{\tIP}, $\pow(\tIP)$, {\tDATA} and {\tMSG}, respectively, and $\msg$ a
data variable of type \tMSG.

Given a valuation of the data variables by concrete data values, the
 process $\cond{\varphi}\p$ acts as $\p$ if $\varphi$
evaluates to {\tt true}, and deadlocks if $\varphi$ evaluates to
{\tt false}.%
\footnote{As
    \label{fn:undefvalues}%
    operators we also allow \emph{partial} functions with the
    convention that any atomic formula containing an undefined subterm
    evaluates to {\tt false}.}
     In case $\varphi$ contains free variables that are not
yet interpreted as data values, values are assigned to these variables
in any way that satisfies $\varphi$, if possible.
The  process $\assignment{\keyw{var}\mathop{:=}\dexp{exp}}\p$
acts as $\p$, but under an updated valuation of the data variables.
The  process $\p+\q$ may act either as $\p$ or as
$\q$, depending on which of the two is able to act at all.  In a
context where both are able to act, it is not specified how the choice
is made. The process
$\broadcast{\dexp{ms}}.\p$ broadcasts (the data value bound to the
expression) $\dexp{ms}$ to all connected {\component}s,
and subsequently acts as $\p$, whereas the process $\unicast{\dexp{dest}}{\dexp{ms}}.\p \prio \q$
tries to unicast the message $\dexp{ms}$ to the destination
\dexp{dest}; if successful it continues to act as $\p$ and otherwise
as $\q$. We abbreviate  $\unicast{\dexp{dest}}{\dexp{ms}}.\p \prio \p$
by $\unicast{\dexp{dest}}{\dexp{ms}}.\p$; this covers the case where
the subsequent behaviour after the unicast is independent of its success.
 The process $\groupcast{\dexp{dests}}{\dexp{ms}}.\p$ tries
to transmit \dexp{ms} to all destinations $\dexp{dests}$, and proceeds
as $\p$ regardless of whether any of the transmissions is successful.
The process $\receive{\msg}.\p$ receives any message $m$ (a
data value of type \tMSG) either from another {\component}, from another
 process running on the same {\component} or from an application layer process
connected to that {\component}.  It then proceeds as $\p$, but with the data
variable $\msg$ bound to the value $m$.  In particular,
$\receive{\newpkt{\dexp{id}}{\dexp{data}}}$
models the injection of
data from the application layer, where the function $\newpktID$ generates
a message containing the application layer
$\dexp{data}$ and the identifier $\dexp{id}$, here indicating the message type.
Data is delivered to the application layer by \deliver{\dexp{data}}.

The internal state of a sequential process described by an expression
$P$ in this language is determined by $P$, together with a
\emph{valuation} $\xi$ associating data values $\xi(\keyw{var})$ to
the data variables \keyw{var} maintained by this process.
In case a process maintains no data values, we use the empty valuation $\xi_0$.
A \emph{valuated process} is a pair $(\xi,P)$ of a sequential process $P$
and an initial valuation $\xi$.

Finally, $V\parl W$ denotes a parallel
composition of valuated processes $V$ and $W$,
with information piped from right to left; in typical applications
\cite{GHPT16} $W$ is a message queue.
This yields a \emph{{\component} expression}.

In the full process algebra \cite{ESOP12}, \emph{node expressions}
$\dval{id}\mathop{:}V\mathop{:}R$ are
given by {\component} expressions $V$, annotated with a component identifier~$\dval{id}$ and
a set of nodes $R$ that are connected to~$\dval{id}$.
In the current application we do not encounter cases were components send
messages to other components that are not connected.
As a consequence, the annotation $\mathop{:}R$ occurring in node expressions is of no significance, and omitted.
We speak instead of \emph{addressed} component expressions $\dval{id}\mathop{:}V$.
In our application $V$ is for example a specification of a generic CAN
driver, whereas $\dval{id}\mathop{:}V$ denotes a specific CAN driver occurring
in the system, such as the one on the mission board.
The identifier $\dval{id}$ of this driver is used as an address by
components that send messages to this driver.

A network is  modelled as a parallel composition of addressed {\component} expressions, using the operator $\|$.

In our specification, we will use a process \keyw{ERROR}, representing
a state that needs to be avoided at all costs. Showing that this state
is in fact unreachable ought to be part of a verification effort.
When specifying a part of a system (as we do in this report)
we make assumptions on the behaviour of components outside our
specification that communicate with our part. The \keyw{ERROR} state
may be reachable if those external components violate the assumptions.
Hence, showing that \keyw{ERROR} is unreachable involves verifying
aspects of the correctness of these external components.\vspace{-3pt}
Formally, we define \keyw{ERROR} as a process name with defining
equation $\keyw{ERROR}() \stackrel{{\it def}}{=} \cond{\false}\,\keyw{ERROR}()$,
representing a \emph{deadlock}. 

% !TEX root = ../paper.tex
\newpage
\section{Formal Specification of all Protocols}\label{app:spec}
The following $6$ sections will present formal specifications of all
the above-mentioned protocols. This includes the specifications themselves, given in \awn, as well as a detailed 
definition of the data structure involved. 
As the semantics of {\awn} is completely unambiguous, specifying a protocol in such a framework enforces total precision and the removal of any ambiguity.

\subsection{Data Structure: Mandatory Types and Messages}\label{sec:datastructure}

In this section we set out the basic data structure needed for the
detailed formal specification of our fragmentation protocol. 
As well as describing
\emph{types} for the information handled at the nodes/components during the
execution of the protocol we also define functions which will be used
to describe the precise intention---and overall effect---of the
various update mechanisms in our protocol.

\subsubsection{Components}

In our formalisation of the CAN bus  we consider a finite
set $\HC$ of \emph{hardware components}; in~\autoref{fig:block} these
are the microcontroller, the mission board, and the camera. 
The defining characteristic of this set is
that there is exactly one CAN driver $\C$ for each hardware component
$H\in\HC$. The COTS component does not count, as it only partakes
to the trusted \can\ via the gateway (cf.\ \autoref{fig:can}).

Furthermore, we consider a finite set $\A$ of \emph{applications}. Each
application is a party that sends messages via the \can. An application
is located on a hardware component, and on each hardware component can be
multiple applications. Figure~\ref{fig:overview} shows two hardware components, each with
two applications.

Finally, there is a finite set $\SC$ of CAN \emph{software components}---the white
rectangles of \autoref{fig:overview}.  Each of them will be specified
by an addressed component expression as defined in \autoref{app:AWN}.
In our model a CAN driver is modelled as the parallel composition of
two software components: one dealing with transmission,
and one with receipt of CAN messages.

\subsubsection{Mandatory Types}
The process algebra \awn always requires the following data structure:
application layer data, 
messages, 
component identifiers
and sets of component identifiers.

\begin{enumerate}
\item The ultimate purpose any communication protocol is to deliver
  \emph{application layer data}.  The type $\tDATA$ describes a
  set of application layer data items. An item of (encrypted) data is thus a particular element of that
  set, denoted by the variables $\data,\ndata\in\tDATA$.
  Since we also inform the application layer about progress, we add special strings such as 
  \msgcomplete\ to the set \tDATA. Moreover, the empty data string is denoted by $\emptystring\in\tDATA$.
\item \emph{Messages} are used to send information via the network. In
  our specification we use the variable $\msg$ of the type $\tMSG$.
  All message types will be described below.
\item
  The type $\tIP$ describes a set of identifiers.
  In our application it is the disjoint union of a set $\tAID$ of \emph{application identifiers}
  (exactly one for each application from $\A$), a set $\tSID$ of \emph{component identifiers}
  (exactly one for each software component from $\SC$) and a set $\tCID$ of CAN IDs.
  Moreover, we use a set $\tMT$ of \emph{message types},
  which we assume to be a subset of $\tCID$.
For each hardware component $H\in\HC$, the constants $\M$, $\R$ and $\C$ of type $\tSID$
  denote the identifiers of the unique multiplexer, reassembly protocol, and transmitting CAN driver within $H$.
  The variable $\sip$ ranges over $\tAID$ and indicates the (ultimate) sender of a message---an application.
  The variables $\keyw{cid}$, $\keyw{bid}$ range over $\tCID$;
  and $\keyw{mt}$ and $\keyw{nmt}$ over $\tMT$.
  Finally we make use of a special element $\mtundef\in\tMT$, denoting the absence of a message.
\end{enumerate}
A message sent over the \can is normally only a fragment of a larger message
(stemming from the application layer), although we allow for CAN messages that are not fragmented.
Henceforth, we use the word \emph{fragment} for such a message.
Message types are allocated to entire messages, whereas
CAN IDs are allocated to fragments. A CAN ID determines the
message type of the whole message, as well as the fragment counter
indicating which fragment of it is currently been transmitted.
A message type uniquely determines the sender of the message, the set of
recipients, and the number of fragments into which the message is
split. Here we take as $\tCID$ an initial segment of the natural numbers.
In our implementation, 11 bits are reserved for CAN IDs.
Given a message type that calls for fragmentation into 3
parts, the CAN IDs form an interval such as 52--54.
The message type is then simply denoted by the first element of
  this interval; in the example 52. It is in this sense that
  $\tMT\subseteq \tCID$.
We use an injective partial function 
\[\begin{array}{r@{\hspace{0.5em}}c@{\hspace{0.5em}}l}
\fncanid : \tMT\times \NN &\rightharpoonup& \tCID  \\
	     \end{array}\]
that, given a message type $\dval{mt}$ and a
fragment counter $k$, which is no larger than the number of fragments
for messages of type $\dval{mt}$, returns a CAN ID\@.
In our example, $\canid{52}{2}=53$.
Due to injectivity, if $\canid{\mt}{\no}$ is defined, the values
$\mt$ and $\no$ can be retrieved.
For the implementation of the protocol, this function is implemented as a static table, 
stored at each component. Sometimes it is possible to reduce the size of this table since 
only those messages types have to be specified that are actually sent or received by 
the fragmentation and reassembly protocols.

\subsubsection{Messages}

Messages are the main ingredient of all our protocols and are used to distribute information. 
The message types used range from new messages to be split and injected by the application layer, via 
messages to acknowledge successful sending, to actual CAN messages that are sent via the {\can}. 
To generate theses messages, we use functions
\[
\begin{array}{r@{\ :\ }l}
\newpktID&\tMT \times \tDATA \rightarrow \tMSG\ ,\\
\fndatamsg & \tMT\times\tDATA\to\tDATA\ ,\\
\canmID & \tCID\times\tDATA \to \tMSG\ ,\\
\cancelID & \tMSG\ ,\\
\canceliID & \tCID\to \tMSG\ ,\\
\ackID & \BB\to \tMSG\ ,\mbox{and}\\
\fnamsg&\tTX\times\tMSG \to \tMSG\ .
\end{array}
\]

A message $\newpkt{\dval{mt}}{\dval{d}}$ is of type \dval{mt} and has the payload $\dval{d}$.
It is injected by the application and received by the fragmentation protocol, which, if necessary, 
then splits the message into smaller fragments.

A reassembled message which is returned to an application by the reassembly protocol is given 
by $\datamsg{\dval{mt}}{\dval{d}}$, where \dval{mt} and \dval{d} are again the message type 
and the actual data, respectively.\footnote{$\fndatamsg$ does not 
generate a message, but is of type $\tDATA$. 
The reason is that, in \awn, messages are used to send information to software components. 
Since we do not model the application, $\fndatamsg$ is a message delivered to the environment, 
which has to be of type $\tDATA$.}

The function $\canm{\dval{cid}}{\dval{d}}$
generates a CAN message with CAN ID \dval{cid}, containing
the data $\dval{d}$.
By $\fragment{\dval{mt}}{\dval{k}}{\dval{d}}$ a fragmented CAN message
is obtained, containing the data $\dval{d}$ and the \mbox{CAN ID} $\canid{\dval{mt}}{\dval{k}}$.
This message is the $\dval{k}^{\rm th}$ fragment of a message of type $\dval{mt}$;
we abstract from all other details of such a CAN message. 

The functions $\cancel{}$ and $\canceli{\dval{cid}}$ are used to
request the cancellation of a message sent before. 
The only difference between these two are that the former tries to
cancel the last message sent, whereas the latter requests the
cancellation of a CAN message with CAN ID \dval{cid}.

The acknowledgement message generated by $\ackM{\dval{b}}$ is
used to communicate the status of a message. $b\in\BB=\{\true,\false\}$ is a Boolean value:
$\ackM{\dval{\true}}$ indicates that the message or fragment was sent successfully, whereas 
$\ackM{\dval{\false}}$ indicates failure, which either stems from a sending problem (e.g.\ a hardware failure) 
or from the message being successfully cancelled. 

The last function we use to generate messages is $\fnamsg$. It is essentially a wrapper function that 
takes an arbitrary message as input and returns the same message with an additional variable attached---
the identifier of a transmit (TX) buffer (an element of the set $\tTX$ of TX buffer identifiers).\footnote{A 
more detailed description of the TX buffers is given in 
\autoref{sec:multiplexer_data2}.}

We assume that all functions building messages are injective, so that
for example the values $\dval{mt}$ and $\dval{data}$ can be retrieved from $\newpkt{\dval{mt}}{\dval{data}}$.
Likewise, we can distil the message type $\dval{mt}$, the fragment number $\dval{no}$ and the payload 
$\dval{d}$ from $\canm{\canid{\dval{mt}}{\dval{no}}}{\dval{d}}$.

\subsubsection{Message Type Table}\label{sec:multiplexer_data1}
Information about the sender, the potential
destinations, the length, and the identifier of the corresponding instance of the
fragmentation protocol for each and every CAN message are stored in a static table.
All components have access to a copy of this table.%
\footnote{Of course, when implementing the protocol, each node would only store the 
bits of the table that are actually needed by the node.}
Such an information table is defined as a set of entries,
exactly one for each message type.

Formally, we define the identifier table as a (total) function
\[
\idtab: \tCID\to \tAID\times \pow(\tSID)\times \NN\times\tSID\ .
\]

Here  $\dval{cid}\mapsto (\dval{aid},\dval{rids}, \dval{parts}, \dval{fid})$
identifies for every CAN ID $\dval{cid}$ (and hence for every message type)
its (unique) sender $\dval{aid} \in\tAID$,
a list of receivers $\dval{rids}\subseteq \tSID$,
the number $\dval{parts}$ of CAN messages
into which the (entire) message needs to be split, and the unique identifier \dval{fid} of the fragmentation protocol handling the CAN message. 
Since we assume that the length of any message of a given type is
fixed, this number can easily be determined.
Note that $\idtab$ is total; hence $\idtab(\dval{cid})$ always returns a value.
We use projections $\pi_{1},\pi_2,\pi_{3},\pi_4$ to select the
corresponding component from the $4$-tuple.

In the formal model (and indeed in any implementation) we need to
extract information form \idtab. To this end, we define the following
functions.
\begin{enumerate}	
\item The sender of a message with CAN ID $\dval{cid}$:
\[\begin{array}{r@{\hspace{0.5em}}c@{\hspace{0.5em}}l}
		    	  \fnsender : \tCID&\to& \tAID\\
	    \sender{\dval{cid}}&:=& \pi_{1}(\idtab(\dval{cid}))
	     \end{array}\]
\item The set of potential (allowed) receivers
of a message with CAN ID $\dval{cid}$:
\[\begin{array}{r@{\hspace{0.5em}}c@{\hspace{0.5em}}l}
		    	  \fnrec : \tCID&\to& \pow(\tSID)\\
	    \rec{\dval{cid}}&:=& \pi_{2}(\idtab(\dval{cid}))
	     \end{array}\]
\item The number of CAN messages into which a message
with CAN ID  $\dval{cid}$ is split:
\[\begin{array}{r@{\hspace{0.5em}}c@{\hspace{0.5em}}l}
		    	  \fnsize : \tCID&\to& \NN\\
	    \size{\dval{cid}}&:=& \pi_{3}(\idtab(\dval{cid}))
	     \end{array}\]
In case no splitting is needed, e.g., if the message is a legacy message, and the message should be handled as
standard CAN message, {\fnsize} should be set to $1$. This guarantees backwards compatibility of our protocol.
\item The name of the instance of the fragmentation protocol responsible for handling the message:
\[\begin{array}{r@{\hspace{0.5em}}c@{\hspace{0.5em}}l}
		\fnfragname:\tCID &\to& \tSID\\
	    \fragname{\dval{cid}}&:=& \pi_{4}(\idtab(\dval{cid}))
	     \end{array}\]
\end{enumerate}
The value $\idtab(\dval{cid})$ only depends on the message type of a
message with CAN ID \dval{cid}; hence
$\idtab(\dval{mt})=\idtab(\canid{\dval{mt}}{k})$ for every $k\leq \size{\dval{mt}}$.

\subsection{The CAN Driver}\label{app:can}

In this section, we present a formal specification of the 
CAN driver, using the process algebra \awn. 
The presented driver is pretty simple and puts 
messages received straight into the corresponding TX buffer. 
In particular it does not provide a message queue to store messages 
until they are handled. Such a queue is not needed since it is part of 
the multiplexer, which we will present later.

\begin{figure}
\centering
\includegraphics[width=7cm]{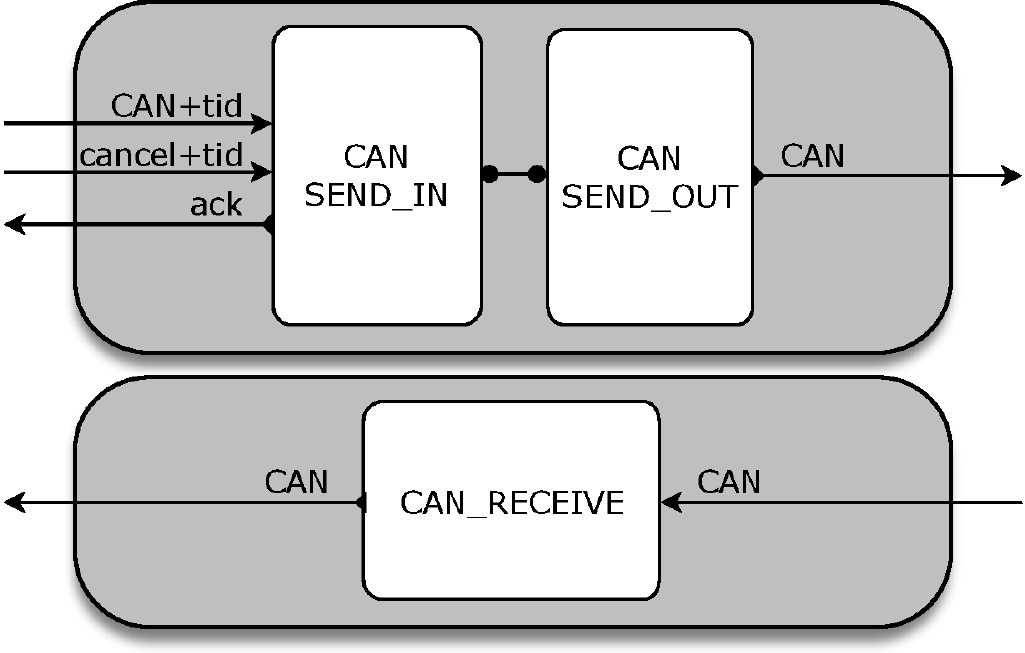}
\caption{Structure of the CAN driver}
\label{fig:messageCan}
\end{figure}

The overall structure of the CAN driver---sketched in 
\autoref{fig:messageCan}---consists of two independent components 
that do not interact. The first handles message sending, the second message receipt. 
\begin{itemize}
\item The protocol that handles message sending is split into two processes: 
the first process, called \CANSIN\, is able to receive messages from the  multiplexer. These messages
are either CAN messages or cancellation messages---both contain a TX-buffer identifier
\dval{tid}. If a CAN message is received, it is stored in the corresponding TX buffer,
and the process \CANSOUT\ is called, which subsumes the behaviour of \CANSIN\ and also handles
the message sending. If a \cancelID-message is received, the process erases the corresponding TX buffer.
\item The protocol that handles message receipt is a single process $\CANR$. It models a simple queue, which stores all incoming messages 
and forwards them to the reassembly protocol as soon as that protocol is ready to handle the next message.
\end{itemize}
All these processes are parametrised with the name $H\mathop\in\HC$
  of the hardware component on which the CAN driver is located.

\subsubsection{Data Structure}\label{sec:can_data}
Each CAN controller provides a number of transmit (TX) buffers. 
Each buffer is able to store a complete CAN message for transmission over the {\can}.
Our architecture offers 3 TX buffers on the microcontroller, and one on the mission board; 
both boards offer 2 receive (RX) buffers.\footnote{The second RX buffer, however,
should not be used due to a hardware bug.}
We assume that every TX buffer on a board has a unique identifier---the set of all identifier is $\tTX$.

We abstract from the concrete number of buffers, and model the buffers
as a (total) function $\fnbuf$ of type $\tTX\to \tMSG$.
If the TX buffer identified by $\dval{tid}$ stores a message $\dval{msg}$, then 
$\buf{\dval{tid}}=\dval{msg}$;
to indicate that the buffer is empty, we use the special element $\msgundef\in\tMSG$.
We define the function space of all these functions as 
\[\begin{array}{r@{\hspace{0.5em}}c@{\hspace{0.5em}}l}
		    	  \BUF \triangleq \tTX&\to& \tMSG\ .
	     \end{array}\]
In our formal specifications we often use an assignment to change a particular value of a function, 
e.g.~$\update{\buf{\txid} := \canm{\cid}{\data}}$ (Line~\ref{cansin:line4} of \Pro{can_send_in}). Implicitly this states that all other values stay the same. 

The CAN driver is supposed to send the message of highest priority (with lowest CAN ID) next. 
To this end we define a partial function 
\[\begin{array}{r@{\hspace{0.5em}}c@{\hspace{0.5em}}l}
		    	  \fnbest : \BUF&\rightharpoonup& \tTX
	     \end{array}\]
that determines the TX buffer that contains this CAN message that is 
most urgent; in case there are different messages with the same priority it chooses non-deterministically.
We claim that
the function \fnbest\ is actually deterministic in our setting.
Formally the function is required to satisfy
\[\begin{array}{cl@{}l}
&\multicolumn{2}{l}{\best{\dval{buffer}} = \dval{tid}}\\
\Leftrightarrow&
\exists \dval{cid}, \dval{d}: \big(&
	\dval{buffer}(\dval{tid}) = \canm{\dval{cid}}{\dval{d}})\ \wedge \\
	&&  (\exists \dval{tid}', \dval{cid}', \dval{d}': \dval{buffer}(\dval{tid}') = \canm{\dval{cid}'}{\dval{d}'} \Rightarrow \dval{cid}\leq \dval{cid}')\big)\ .
\end{array}\]
Note that $\best{\dval{buffer}}$ is undefined if and only if all TX buffers are empty, i.e.,
if $\dval{buffer}(\dval{tid})=\msgundef$ for all $\dval{tid}\in \tTX$.

We use a queue-style data structure for modelling an inbox of the
CAN receiver. In general, we denote queues of messages by
$[\tMSG]$, denote the empty queue by $[\,]$, and make use of the
standard (partial) functions
$\fnhead:[\tMSG]\rightharpoonup\tMSG$,
$\fntail:[\tMSG]\rightharpoonup[\tMSG]$ and
$\fnappend:\tMSG\times[\tMSG]\rightarrow[\tMSG]$ that return the
``oldest'' element in the queue, remove the ``oldest'' element, and
add a packet to the queue, respectively. 

In our protocol specification below, all messages received via the {\can} are stored until the reassembly 
protocol is ready to handle them. 
This may not be needed in the implementation of this protocol, as a timing analysis
may show that when a message arrives, the protocol is always ready to handle it.
However, in the forthcoming verification of the correctness of our protocol we do not want to depend
on this timing analysis, and hence incorporate incoming message queues. Thus, a separate verification---involving
the timing analysis---will be needed to show that these queues can be omitted.

At the moment we assume an infinite queue, which is unrealistic. 
It is an easy task to model a finite queue, where messages are lost in case the buffer is full. 
Of course then properties such as ``every message sent will be received'' may not hold; 
they require carefully designed preconditions.

This section concludes with a table summarising 
the entire data structure we use for the CAN driver.
It summarises not only the data structure presented in this section, but 
also recapitulates the necessary part of the structure presented in \autoref{sec:datastructure}.
Similar tables will be given at the end of every section that discusses data structure
(cf. Sections~\ref{sec:frag_data}, \ref{sec:reass_data} and \ref{sec:multiplexer_data2}).

{\centering{\small
\setlength{\tabcolsep}{4.0pt}
\begin{longtable}{@{}|l|l|l|@{}}
\hline
\textbf{Basic Type} & \textbf{Variables} & \textbf{Description}\\
\hline
 \tMSG		&\msg	&messages\\
 \tDATA		&\data	&data/payload of messages\\
 \tTX		&\txid	&identifiers for TX buffers\\
 \tCID 		&\cid	& CAN IDs\\
 \tSID		&	&CAN software component identifiers\\ 
\hline
\hline
\textbf{Complex Type} & \textbf{Variables} & \textbf{Description}\\
\hline
$\BUF \triangleq \tTX\to \tMSG$ &\fnbuf & array of (CAN) messages, modelling the TX buffers\\
${[\tMSG]}$						&\msgs			&message queues\\
\hline
\hline
\multicolumn{2}{|l|}{\textbf{Constant}}& \textbf{Description}\\
\hline
\multicolumn{2}{|l|}{$\msgundef:\tMSG$}&special message symbol (indicating absence of a message)\\
\multicolumn{2}{|l|}{${[\,]}:{[\tMSG]}$}&	empty queue\\
\multicolumn{2}{|l|}{$\M: \tSID$}& multiplexer identifier for hardware component $H$\\
\multicolumn{2}{|l|}{$\R: \tSID$}& reassembling protocol identifier for hardware component $H$\\
\multicolumn{2}{|l|}{$\C: \tSID$}& transmitting CAN driver identifier for hardware component $H$\\
\multicolumn{2}{|l|}{$\D: \tSID$}& receiving CAN driver identifier for hardware component $H$\\
\hline
\hline
\multicolumn{2}{|l|}{\textbf{Function}} & \textbf{Description}\\
\hline
\multicolumn{2}{|l|}{$\cancelID: \tMSG$}&
	cancellation message\\
\multicolumn{2}{|l|}{$\ackID : \BB\to \tMSG$}&
	acknowledgement to multiplexer\\
\multicolumn{2}{|l|}{$\canmID : \tCID\times\tDATA \to \tMSG$}&
	create CAN messages out of identifier and payload\\
\multicolumn{2}{|l|}{$\fnamsg:\tTX\times\tMSG \to \tMSG$}&
	wrapper function to add a TX-identifier to a message\\
\multicolumn{2}{|l|}{$\fnrec : \tCID\to \pow(\tSID)$}&
	 set of potential (allowed) receivers of a message\\
\multicolumn{2}{|l|}{$\fnbest : \BUF\rightharpoonup \tTX$}&
	 returns the name of the TX that contains the `best' CAN message\\	 	 
\multicolumn{2}{|l|}{$\fnhead:[\tMSG]\rightharpoonup\tMSG$}&
	returns the `oldest' element in the queue\\
\multicolumn{2}{|l|}{$\fntail:[\tMSG]\rightharpoonup[\tMSG]$}&
	removes the `oldest' element in the queue\\
\multicolumn{2}{|l|}{$\fnappend:\tMSG\times[\tMSG]\rightarrow[\tMSG]$}&
	inserts a new element into the queue\\
\hline
\caption[]{\rule[10pt]{0pt}{3pt}Data structure for the CAN Controller/Driver}
\end{longtable}
}}

\subsubsection{Formal Specification}\label{sec:can_spec}

\myparagraph{Sending Messages} 
The sending procedure consists of two different processes.
The first solely deals with receiving messages (from the multiplexer); and the second also with
sending messages on the \can. The second process subsumes the behaviour of the first process by
offering it as a alternative to sending. The first process is only called in the initial state of
the protocol, and as a potential behaviour of the second process.
Both processes maintain a single variable $\fnbuf$, which models the corresponding TX buffer (see above).

  \algsetup{linenodelimiter=.,linenosize=\tiny}
  \begin{algorithm}[ht]
    {\footnotesize
      \caption{CAN driver---Sending Routine I}
      \label{pro:can_send_in}
      \begin{algorithmic}[1]
        % !TEX root = ../paper.tex
\DEFPROCESS{\CANSIN}{\fnbuf}																												\label{cansin:line0}
		\receiveL{\msg}\ . 																																\label{cansin:line1}
   		\PAR                         																															\label{cansin:line2}
			 \IF[new CAN message for TX buffer \txid]{$\msg = \amsg{\txid}{\canm{\cid}{\data}}$}     				\label{cansin:line3}
				\UPD{\buf{\txid} := \canm{\cid}{\data}} \COM{override TX buffer}													\label{cansin:line4}													
				\cansoutL{\fnbuf}																														\label{cansin:line5}
			\ELSIF[cancellation message for TX buffer \txid]{$\msg = \amsg{\txid}{\cancel{}}$} 						\label{cansin:line6}
	        		\PAR																																			\label{cansin:line7}
				\IF[TX buffer {\txid} is already cleared]{$\buf{\txid} = \msgundef$}     											\label{cansin:line8}
					\cansoutL{\fnbuf}																													\label{cansin:line9}
				\ELSIF[TX buffer {\txid} contains a message]{$\buf{\txid} \neq \msgundef$} 								\label{cansin:line10}
             		\UPD{\buf{\txid} := \msgundef}\COM{erase TX buffer}															\label{cansin:line11}
%		\COMLINE{in Phase 3 we have to show that we cannot deadlock here, i.e., that the multiplexer is not sending at the same time}
					\unicastL{\M}{\amsg{\txid}{\ackM{\false}}}\ . \									\label{cansin:line12}
					\cansoutL{\fnbuf}																													\label{cansin:line13}
       			 \ENDIFii					
        			\ENDPAR																																	\label{cansin:line14}											
			\ENDIFii
		\ENDPAR																																			\label{cansin:line15}		

	\end{algorithmic}
    }%end{footnotesize}
  \end{algorithm}

The first process, named \CANSIN\ and depicted in \Pro{can_send_in}, starts with receiving a message $\msg$ from the 
connected multiplexer $\M$ (Line~\ref{cansin:line1}).
After that the process checks the type of the message received:

In case the message is a CAN message with an additional value $\txid$, which indicates the 
destined TX buffer, the process stores the message into the TX buffer $\txid$ (Line~\ref{cansin:line4}) and 
calls the process \CANSOUT, which handles the sending of CAN
messages, as well as the processing of further messages from the multiplexer (by calling \CANSIN).
Note that the new message is copied into the TX buffer regardless whether the buffer already contains a 
message. If it does, that previous message is lost. It is up to the multiplexer (\autoref{sec:multiplexer}) to avoid such 
a scenario.

In case the process receives a cancellation request for TX buffer $\txid$ (the incoming message 
has the form $\amsg{\txid}{\cancel{}}$, Line~\ref{cansin:line6}), depending on the status 
of the corresponding TX buffer, different actions are performed. 
If buffer $\txid$ is empty (Line~\ref{cansin:line8}) there is no message to be cancelled and the 
protocol performs no action; it calls the process \CANSOUT\ to proceed.
If the buffer is not empty (Line~\ref{cansin:line10}) the process erases the buffer.
After that the multiplexer is informed about the successful cancellation, which is done by 
unicasting the message $\amsg{\txid}{\ackM{\false}}$ to the connected multiplexer \M.
Since our unicast is blocking (the protocol is stuck if the recipient of the message is not ready to receive the message),
there could be a possibility that the protocol deadlocks while trying to send the acknowledgement.
To prevent this, we make sure that the multiplexer is \emph{input
    enabled}, meaning that the multiplexer is always ready to receive messages (cf.~\autoref{sec:multiplexer}).

In case another type of message would be received the protocol deadlocks after message receipt.
However, we can show that no other message type can be
received and hence the protocol is free of deadlocks.

The main purpose of the second process, named \CANSOUT\ and depicted in \Pro{can_send_out}, 
is to send the CAN messages stored in the TX buffers. It also offers
the behaviour of \CANSIN, to process another incoming message as an alternative to any activity that could block further progress.

  \algsetup{linenodelimiter=.,linenosize=\tiny}
  \begin{algorithm}[ht]
    {\footnotesize
      \caption{CAN driver---Sending Routine II}
      \label{pro:can_send_out}
      \begin{algorithmic}[1]
        % !TEX root = ../paper.tex

\DEFPROCESS{\CANSOUT}{\fnbuf}																														\label{cansout:line0}
	\IFempty
		\cansinL{\fnbuf}\COM{another message incoming}																							\label{cansout:line1}
	\ELSIF[messages in TX buffers to be sent]{$\txid = \best{\fnbuf}\wedge \buf{\txid} = \canm{\cid}{\data}$}	\label{cansout:line2}
		\PAR																																								\label{cansout:line3}
		\cansinL{\fnbuf}																																				\label{cansout:line4}
		\STATE$+$																																					\label{cansout:line5}
	     	\groupcastL{\rec{\cid}}{\canm{\cid}{\data}}\ .					\label{cansout:line6}							
     	\UPD{\buf{\txid} := \msgundef}																														\label{cansout:line7}
     	\unicastL{\M}{\amsg{\txid}{\ackM{\true}}}\ .															\label{cansout:line8}
		\cansoutL{\fnbuf}																																			\label{cansout:line9}
		\ENDPAR																																						\label{cansout:line10}
	\ENDIFii

	\end{algorithmic}
    }%end{footnotesize}
  \end{algorithm}

If there is another incoming message, the process may run \CANSIN\ (Line~\ref{cansout:line1}); 
otherwise \CANSOUT\ determines the CAN message that should be send next (if any). This 
is done by use of the function $\fnbest$ in Line~\ref{cansout:line2}. 
Since \fnbest\ returns the name of a TX buffer that contains a CAN message\footnote{In fact
an invariant we might need to show is that all TX buffers only contain CAN messages.}
the second conjoint $\buf{\txid} = \canm{\cid}{\data}$ is not a restriction---it is used to distil 
the CAN ID $\cid$ of that message.

In Line~\ref{cansout:line6} the process 
sends the CAN message $\canm{\cid}{\data}$ with highest priority to
the intended recipients of this message, which are determined by the function $\fnrec$.
This transmission uses the CAN bus and is carried out by the CAN controller.
It will succeed only when there are no CAN messages with higher
priority sent by other CAN controllers on the bus.
As long as the transmission is pending, the process has the option to
process another incoming message by executing \CANSIN\ (Line~\ref{cansout:line4}).
 After the message has been successfully sent via the {\can}, the TX buffer is erased 
(Line~\ref{cansout:line7}) and the connected multiplexer is informed about success by an \ackID-message (Line~\ref{cansout:line8}).
As remarked before, the multiplexer is always ready to accept this acknowledgement.
Finally, the process returns to \CANSOUT\ to either accept another incoming message from the
multiplexer or to transmit another buffered message on the \can.

% In the operational semantics of AWN \cite{ESOP12} the evaluation of a guard, as in
% Line~\ref{cansout:line2}, induces a state change. For this reason a process
% $P + [\phi](P+Q)$ is not equivalent to $P + [\phi]Q$. Leaving out Line~\ref{cansout:line4}
% of \Pro{can_send_out} can give rise to stagnation. It can occur that the \textbf{groupcast} of Line~\ref{cansout:line6},
% by which the CAN driver sends a message on the \can, is blocked as long as the \can is busy
% sending packets with higher priority from other nodes. When this happens, the CAN driver should
% still be able to receive input from the multiplexer.
Line~\ref{cansout:line4} guarantees that the process can receive another message 
(via Line~\ref{cansin:line1} of \Pro{can_send_in}). Without this line \Pro{can_send_out} can
give rise to stagnation (deadlock), since the \textbf{groupcast}-action of Line~\ref{cansout:line6}
can only succeed if \tcan is not busy sending higher-priority messages from other nodes.

\myparagraph{Receiving Messages}
We assume that any message sent via the {\can} 
is actually received by all CAN controllers/drivers that are supposed 
to receive the message.  For this reason, a CAN driver should
always be able to perform a receive action, regardless of which state
it is in. We introduce a process $\CANR$ (see \Pro{can_receive}), modelling a message queue,
that runs in parallel with {\CANSIN/ \CANSOUT}.
Every incoming message is stored in this queue, and piped from
there to the reassembly protocol {\RASS}, which we will discuss later in \autoref{sec:reass}, whenever {\RASS} is ready to handle a
new message. The process {\CANR} is always ready to receive a new
message. 

Similar to the process call in \Pro{can_send_out} the {\bf receive}-action in Line~\ref{canr:line8} is needed 
to guarantee that messages can be received at any time.

  \algsetup{linenodelimiter=.,linenosize=\tiny}
  \begin{algorithm}[ht]
    {\footnotesize
      \caption{CAN driver---Receiving Routine}
      \label{pro:can_receive}
      \begin{algorithmic}[1]
        % !TEX root = ../paper.tex
\DEFPROCESS{\CANR}{\msgs}
	\IFempty
		\COMLINE{store incoming message at the end of \msgs}		
		\receiveL{\msg}\ . 			
		\canr{\append{\msg}{\msgs}}					
	\ELSIF[the queue is not empty]{$\msgs\not=[\,]$}			
		\PAR
		\COMLINE{pop top message and send it to the reassembly protocol}		
\unicastL{\R}{\head{\msgs}}\ .\ 
		\canr{\tail{\msgs}}												
		\COMLINE{or receive and store an incoming message}				
		\STATE $+$\,  \receive{\msg}\ . \canr{\append{\msg}{\msgs}}        \label{canr:line8} 
		\ENDPAR
	\ENDIFii

	\end{algorithmic}
    }%end{footnotesize}
  \end{algorithm}

\subsubsection{Initialisation}\label{ssec:can_init}
To finish our specification, we have to define an initial state for the CAN driver. 
The initial state requires the assignment of any variable occurring in a process. 
Such an assignment is provided by  \emph{valuation functions},
offered by \awn\ (cf.~\autoref{app:AWN}).

The initial process $C_H$ of the CAN driver on hardware
component $H\in\HC$ is given by the expression
\[
\C\mathop:(\xi,\cansin{\fnbuf})\ \| \D \mathop{:}(\xii,\canr{\msgs})\ ,
\]
with $\C$ and $\D$ the component identifiers of the sending
and receiving CAN components, and with
\begin{equation*}
\xi(\buf{\txid})=\dval{\msgundef} \quad (\forall \txid\in\tTX)
\ \wedge\
\xii(\msgs)=[\,]\ .
\end{equation*}
This says that initially all TX buffers controlled by a CAN driver are empty,  
and the message queue of CAN messages received by a CAN receiver is empty.

\subsection{Fragmentation Protocol}\label{sec:frag}

\begin{figure}
\centering
\includegraphics[width=6.8cm]{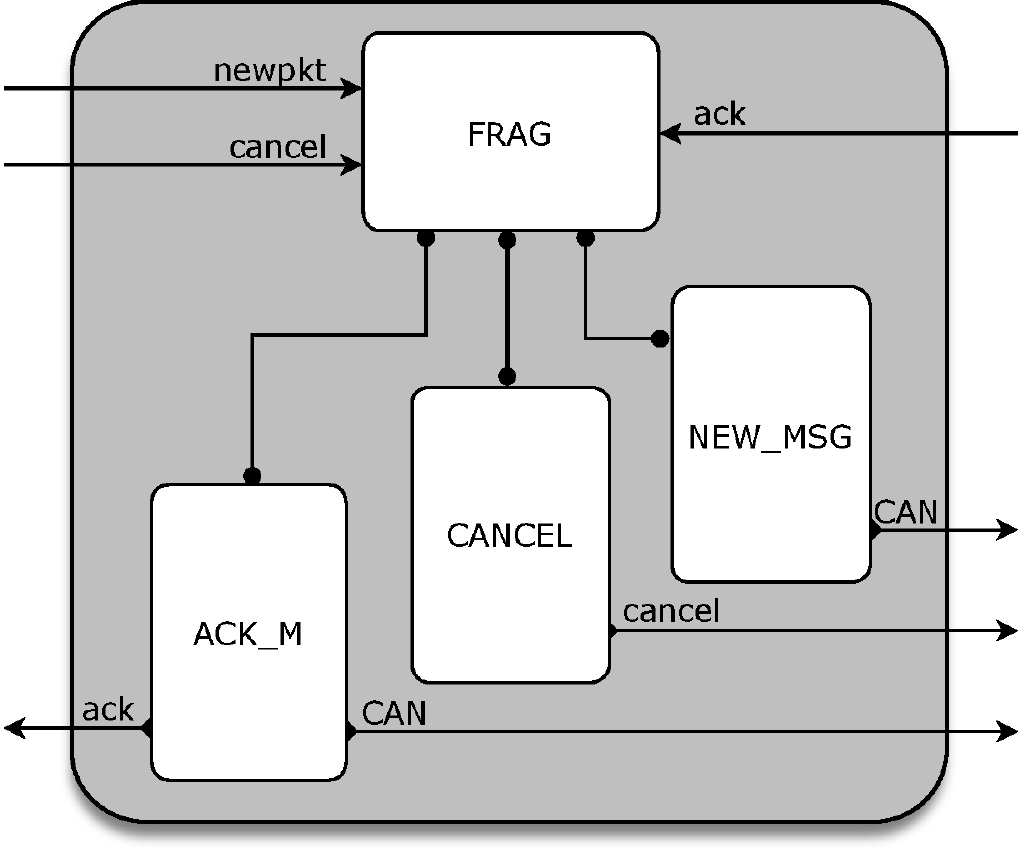}
\caption{Structure of the Fragmentation Protocol}
\vspace{-2mm}
\label{fig:message_frag}
\end{figure}

In this section, we present a formal specification of the fragmentation
protocol. 
Our model, which is sketched in \autoref{fig:message_frag}, consists of 4 processes, named \FRAG, \MSG, \CANC\ and \ACKM:
\begin{itemize}
\item The basic process \FRAG\ receives a message from the application layer or from the multiplexer and, depending on the type of the message, calls other processes. When there is no message handling going on, it idles until a new message arrives.
\item The process \MSG\ describes all actions performed by the
  fragmentation protocol when a new message is received from the
  application. This includes the possibility of reaching an
  unrecoverable error state in case the application injects a message
  before a previous message was successfully handled or cancelled.
  The application layer should be programmed in such a way that this
  error state is always avoided.  
\item The process \CANC\ handles all actions to be performed when an application request the cancellation of a message previously sent.
\item The process \ACKM\ describes the protocol behaviour in case an
  acknowledgement message (positive or negative) is received from the multiplexer.  
  Depending on the situation, this process reports the status to the
  application, sends the next fragment, or reaches an unrecoverable
  error state.
\end{itemize}

\subsubsection{Data Structure}\label{sec:frag_data}

The main intention of the protocol is to split  data given from the application layer. 
For this purpose we define functions to manipulate data. 
Since all these functions can be seen as bit-level operations, we are not giving the exact definitions.

The function $\fnheaddata:\tDATA\to\tDATA$ extracts the first $8$ bytes from a given \data; or returns the entire data in case it is shorter.

The function $\fntaildata:\tDATA\to\tDATA$ is the complement of
  $\fnheaddata$ and returns the remaining data (if any) after $8$ bytes have
  been chopped off.

The following table summarises the entire data structure we use for the fragmentation protocol.
{\centering{\small
\setlength{\tabcolsep}{4.0pt}
\begin{longtable}{@{}|l|l|l|@{}}
\hline
    \textbf{Basic Type} & \textbf{Variables} & \textbf{Description}\\
\hline
    \tMSG		&\msg	&messages\\
    \tDATA		&\data, \ndata	&stored data\\
    \NN		&\no		&fragment counter\\
    \BB 		&\abort, \ackb 	&Boolean flags\\
    \tMT		&\mt, \nmt	&message types\\
    \tCID 	&		& CAN IDs\\
    \tSID		&	&CAN software component identifiers\\ 
\hline
\hline
\multicolumn{2}{|l|}{\textbf{Constant}}& \textbf{Description}\\
\hline
\multicolumn{2}{|l|}{$\emptystring:\tDATA$}&
	the empty data string\\
\multicolumn{2}{|l|}{$\mtundef:\tMT$}&
	special message type symbol, denoting undefined message type\\
\multicolumn{2}{|l|}{$\M: \tSID$}& multiplexer identifier for hardware component $H$\\
\hline
\hline
\multicolumn{2}{|l|}{\textbf{Function}} & \textbf{Description}\\
\hline
\multicolumn{2}{|l|}{$\newpktID: \tMT\times \tDATA \to\tMSG$}&
	creates application-layer message out of message type and data\\
\multicolumn{2}{|l|}{$\cancelID: \tMSG$}&
	cancellation message from application layer\\
\multicolumn{2}{|l|}{$\canceliID : \tCID \to \tMSG$} & 
	cancellation message to multiplexer\\
\multicolumn{2}{|l|}{$\ackID : \BB\to \tMSG$}&
	acknowledgement from multiplexer\\
\multicolumn{2}{|l|}{$\canmID : \tCID\times\tDATA \to \tMSG$}&
	create CAN messages out of identifier and payload\\
\multicolumn{2}{|l|}{$\fncanid: \tMT\times\NN \rightharpoonup \tCID$}&
	returns a CAN ID for a given message type and a fragment counter\\
\multicolumn{2}{|l|}{$\fnheaddata : \tDATA\to \tDATA$}&
	takes the first $8$ bytes from a given data streams\\
\multicolumn{2}{|l|}{$\fntaildata : \tDATA\to \tDATA$}&
	removes the first $8$ bytes from a given data streams\\
\multicolumn{2}{|l|}{$\fnsize : \tCID\to \NN$}&
	number of CAN messages into which a message is split\\
\hline
\caption[]{\rule[10pt]{0pt}{3pt}Data structure for the Fragmentation Protocol}
\end{longtable}
}}

\subsubsection{Formal Specification}\label{sec:frag_spec}

\myparagraph{The Main Loop}
The basic process \FRAG\ receives messages from the application layer or the multiplexer; one at a time. 
This process maintains four data variables \mt, \data, \no, and \abort, in which it stores 
the message type last handled, 
the data received from the application and not yet fragmented, 
the number of fragments already sent, 
and a flag to signal that the application requested cancellation.
The process is considered to be currently handling an
application-layer message iff the value of $\no$ is non-zero.

  \algsetup{linenodelimiter=.,linenosize=\tiny}
  \begin{algorithm}[ht]
    {\footnotesize
      \caption{Fragmentation---Main Loop}
      \label{pro:frag}
      \begin{algorithmic}[1]
        % !TEX root = ../paper.tex
\DEFPROCESS{\FRAG}{\mt\comma\data\comma\no\comma\abort}										\label{frag:line0}
    \receiveL{\msg}\ .                                  					                                    				      		\label{frag:line1}
    \PAR                         																											\label{frag:line2}
    \IF[new message to be sent; distill {\nmt} and {\ndata}]{$\msg = \newpkt{\nmt}{\ndata}$}   	\label{frag:line3}
		\nmsgL{\nmt}{\ndata}{\no}{\abort}																					\label{frag:line4}
	\ELSIF[cancellation message received]{$\msg = \cancel$} 												\label{frag:line5}
		\cancL{\mt}{\data}{\no}{\abort}																						\label{frag:line6}
	\ELSIF[message from multiplexer]{$\msg = \ackM{\ackb}$} 												\label{frag:line7}
		\ackmL{\ackb}{\mt}{\data}{\no}{\abort}																			\label{frag:line8}
    \ENDIFii
	\ENDPAR																															\label{frag:line9}	
	\end{algorithmic}
    }%end{footnotesize}
  \end{algorithm}

The routine is formalised in \Pro{frag}.
First, a message has to be received using the command \receive{\msg} (Line~\ref{frag:line1}). 
The message stems either from the application or from the multiplexer (cf.\ \autoref{sec:overview}).
After that, the process \FRAG\ checks the type of the message and calls a process that can handle this message: 
in case of a `fresh' message from the application layer, the process
\MSG\ is called (Line~\ref{frag:line4})---note that during the process
call the values of $\mt$ and $\data$ are updated with the
new values $\nmt$ and $\ndata$ extracted
from the message;  
in case of an incoming cancellation request the process \CANC\ is executed (Line~\ref{frag:line6});
and in case a message from the multiplexer ($\ackID$) is read, the process \ACKM\ is called,
carrying the Boolean flag {\ackb} received from the multiplexer.
In case a message of any other type is received, the process
\FRAG\ deadlocks; it is a proof obligation to check that this will not occur.

\myparagraph{New Message}
The process \MSG\ describes all actions performed by the fragmentation protocol when a new message is injected by 
an application hooked up to the instance of the fragmentation protocol. 

  \algsetup{linenodelimiter=.,linenosize=\tiny}
  \begin{algorithm}[ht]
    {\footnotesize
      \caption{New Message Received}
      \label{pro:msg}
      \begin{algorithmic}[1]
        % !TEX root = ../paper.tex
\DEFPROCESS{\MSG}{\mt\comma\data\comma\no\comma\abort}        			\label{msg:line0}                                                                                                   
    \IF[protocol is sending another message]{$\no\not=0$} 						\label{msg:line1}
     \errL  \COM{Enter error state and die}									\label{msg:line2}
	\ELSIF[protocol ready to start handling new message]{$\no=0$} 			\label{msg:line3}
	 	\unicastL{\M}{\fragment{\mt}{1}{\headdata{\data}}}\ .\ 
		\frag{\mt}{\taildata{\data}}{1}{\abort}								\label{msg:line4}
    \ENDIFii

	\end{algorithmic}
    }%end{footnotesize}
  \end{algorithm}

In case the fragmentation protocol is in the process of sending (fragmenting) a message, which was previously received, 
the value of $\no$ is different to $0$, since this integer counts the number of fragments already sent.
If this is the case (Line~\ref{msg:line1}), the process deadlocks with an error; the entire fragmentation stops and cannot be restored. 
It is up to the application hooked up to the fragmentation protocol to avoid this scenario. 

In case the fragmentation protocol is not handling another message ($\no = 0$), the process \MSG\ starts splitting the data:
it chops off the first $8$ bytes of \data, using the function $\fnheaddata$, and then creates a CAN message, which is sent to 
the connected multiplexer $\M$. The corresponding CAN ID is determined by $\canid{\mt}{1}$---the first fragment of message type {\mt} is being 
handled. Before the second fragment can be sent to the multiplexer, the fragmentation protocol has to await an acknowledgement of the 
multiplexer (which actually stems from the CAN driver and is only forwarded by the multiplexer). Hence the process returns to \FRAG,
where it can receive a new message, while updating the fragment counter $\no$ and the remaining data to be fragmented.

\myparagraph{Cancellation Message} An application always has the possibility to cancel 
the message previously sent. Since one of our assumptions is that the application layer only submits one message a time 
(otherwise the fragmentation protocol yields a deadlock---see Line~\ref{msg:line2} of \Pro{msg}), only the last message 
can be cancelled.

  \algsetup{linenodelimiter=.,linenosize=\tiny}
  \begin{algorithm}[ht]
    {\footnotesize
      \caption{Cancellation Request}
      \label{pro:canc}
      \begin{algorithmic}[1]
        % !TEX root = ../paper.tex
\DEFPROCESS{\CANC}{\mt\comma\data\comma\no\comma\abort}				\label{canc:line0}
    \IF[protocol is sending a message]{$\no\not=0$} 										\label{canc:line1}
        \unicastL{\M}{\canceli{\canid{\mt}{\no}}}\ .\ \frag{\mt}{\data}{\no}{\true}		\label{canc:line2}
	\ELSIF[protocol is not sending a message]{$\no=0$} 								\label{canc:line3}
	 	\fragL{\mt}{\data}{\no}{\abort}\COM{nothing to cancel}							\label{canc:line4}
    \ENDIFii

	\end{algorithmic}
    }%end{footnotesize}
  \end{algorithm}

Similar to \Pro{msg} the cancellation process checks whether the fragmentation protocol is currently sending/splitting a message; as before this check is done by evaluating the statement 
$\no = 0$. In case the protocol is currently working on splitting a message, the cancellation request is forwarded to the multiplexer {\M} (Line~\ref{canc:line2}); 
it is the multiplexer who can stop the current fragment to be sent.
Since the multiplexer is connected to multiple instances of the fragmentation protocol, the cancellation message must now carry the unique CAN ID of the message to be cancelled.
 The fragmentation protocol itself just stops splitting the {\data} further; it signals the cancellation process
 by setting the flag {\abort} to {\true}. The protocol does not yet set the value of $\no$ to $0$---indicating that it is idle again. This is only done after an acknowledgement from the multiplexer has been received. 
In case the fragmentation protocol is idle and receives a cancellation message---this can happen in case that the cancellation message was sent by the application just before the acknowledgement of successful transmission was sent to the application---the cancellation request is ignored and the protocol returns to the Process \FRAG\ (Line~\ref{canc:line4}).

\myparagraph{Notification from the Multiplexer}
Independent of whether a transmission was successful or not, the CAN controller will send an acknowledgement via the 
multiplexer: in case of a successful transmission the contents of this message is `\true', in case of a failure, or a cancellation, the value `\false' is transmitted---the value is stored temporarily in the variable \ackb.

  \algsetup{linenodelimiter=.,linenosize=\tiny}
  \begin{algorithm}[ht]
    {\footnotesize
      \caption{Acknowledgement from the Multiplexer}
      \label{pro:ackmult}
      \begin{algorithmic}[1]
        % !TEX root = ../paper.tex
\DEFPROCESS{\ACKM}{\ackb\comma\mt\comma\data\comma\no\comma\abort}									\label{ackmult:line0}
	\IF[process sending]{$\no\not=0$} 																									\label{ackmult:line1}
	\PAR																																					\label{ackmult:line2}
    		\IF[positive acknowledgment received ]{$\ackb=\true$} 																	\label{ackmult:line3}
    		\PAR																																				\label{ackmult:line4}
	         \IF[last fragment sent successfully]{$\no=\size{\mt}$} 																\label{ackmult:line5}
	         	\deliverL{\msgcomplete}\ .\ \frag{\mt}{\varepsilon}{0}{\false}													\label{ackmult:line6}
			\ELSIF[send next fragment]{$\no\not=\size{\mt} \wedge \abort=\false$} 										\label{ackmult:line7}
				\unicastL{\M}{\fragment{\mt}{\no+1}{\headdata{\data}}}\ .\ \frag{\mt}{\taildata{\data}}{\no+1}{\false}\!\!\!		\label{ackmult:line8}
			\ELSIF[message cancelled, no fragment to be sent]{$\no\not=\size{\mt} \wedge \abort=\true$} 	\label{ackmult:line9}
				\deliverL{\msgfailed}\ .\ \frag{\mt}{\varepsilon}{0}{\false}															\label{ackmult:line10}
	         \ENDIFii
    		\ENDPAR																																		\label{ackmult:line11}
		\ELSIF[sending failed or aborted]{$\ackb=\false$} 																			\label{ackmult:line12}
			\deliverL{\msgfailed}\ .\ \frag{\mt}{\varepsilon}{0}{\false}																\label{ackmult:line13}
		\ENDIFii							
	\ENDPAR																																			\label{ackmult:line14}
	\ELSIF[process is not sending]{$\no=0$} 																							\label{ackmult:line15}
	\PAR																																					\label{ackmult:line16}
		\IF[positive acknowledgment received ]{$\ackb=\true$}																	\label{ackmult:line17} 
			\errL																																			\label{ackmult:line18}
		\ELSIF[sending failed or aborted]{$\ackb=\false$} 																			\label{ackmult:line19}
			\fragL{\mt}{\data}{\no}{\abort}																										\label{ackmult:line20}
		\ENDIFii
	\ENDPAR																																			\label{ackmult:line21}
	\ENDIFii

	\end{algorithmic}
    }%end{footnotesize}
  \end{algorithm}

As in the previous process, \Pro{ackmult} only takes actions when $\no \not= 0$, meaning
that the process is in a state where it handles a message.
In case the process is not handling a message (Line~\ref{ackmult:line15}) it either deadlocks
with an error (in case of a positive acknowledgement---Lines~\ref{ackmult:line17}--\ref{ackmult:line18}),
or ignores the message (Line~\ref{ackmult:line20}). In the future we hope to show that both situations
cannot occur. The asymmetric treatment is immaterial, and follows the implementation.

Lines~\ref{ackmult:line1}--\ref{ackmult:line14} handle the case when the process receives an acknowledgement and is handling a message right now---the standard case.
Lines~\ref{ackmult:line4}--\ref{ackmult:line11} handle the case of a positive acknowledgement, indicating the transmission of the last fragment to the multiplexer---and thereby to the CAN controller---was successful. If this is the case, the fragmentation protocol uses the function $\fnsize$ to determine whether the last fragment of the split message was sent. If this is the case (Line~\ref{ackmult:line5}), the protocol informs the application about the success ($\deliver{\msgcomplete}$), sets the values of $\no$ and $\abort$ to $0$ and $\false$, respectively (to indicate that the process is ready to receive a new message from the application), and returns to the main process \FRAG. 
In case the message that is currently handled has not been sent entirely and was not aborted (no cancellation request received), the protocol 
chops the next $8$ bytes of \data\ to be sent and creates a new CAN message, which is passed on to the multiplexer \M\ (Line~\ref{ackmult:line8});
the local data ($\no$ and $\data$) are adapted accordingly.
In case a positive acknowledgement is received, but the message was cancelled meanwhile (Line~\ref{ackmult:line9}), 
the protocol informs the application layer and returns to an idle state, with $\no$ set to $0$, the $\data$ being `removed' and the Boolean flag \abort\ set to \false\ 
(Line~\ref{ackmult:line9}).
In case the fragmentation protocol receives a negative acknowledgement, the protocol behaves the same: it informs the application layer, and returns to an idle state.

\subsubsection{Initialisation}\label{ssec:frag_init}

The fragmentation protocol $F_{\dval{id}}$ is initialised by 
$\dval{id}\mathop:(\xi,\frag{\mt}{\data}{\no}{\abort})$, 
with \dval{id} the identifier of any particular instance of the fragmentation protocol, and
\begin{equation*}
\xi(\mt) = \mtundef
\ \wedge\
\xi(\data) = \emptystring
\ \wedge\
\xi(\no) = 0
\ \wedge\
\xi(\abort) = \false\ .
\end{equation*}
This says that no message has been received, and no data is stored; 
moreover the protocol is neither sending nor aborting a message.

\subsection{Reassembly Protocol}\label{sec:reass}
The reassembly protocol collects fragments of messages sent via the {\can}. 
It stores  and reassembles the messages; as soon as a message is fully received, 
the message is sent to the application layer. The protocol also checks whether 
some fragments are lost and drops partly assembled messages that cannot be completed. It can be implemented in AWN as a single process: 
\RASS.  
  
\subsubsection{Data Structure}\label{sec:reass_data}

If a fragment of a message is received by a component, 
it may need to be stored until the full message can be recreated by reassembling the fragments. 
Since reordering does not happen on the \can, and using assumption (7) in \autoref{sec:assum},
it suffices to have a single data storage for 
every application as sender of the fragment.\footnote{This
    differs from an earlier specification and implementation. There we assumed a data storage for every message type; the
new version improves memory storage dramatically.}

Hence, the data type\vspace{-1ex} of the store is defined as the function space
\[\begin{array}{r@{\hspace{0.5em}}c@{\hspace{0.5em}}l}	
\tQUEUES \triangleq \tAID &\to&  \tMT \times \NN \times\; \tDATA\;.\\
\end{array}\]%
For every sender (an application, with unique ID $\dval{aip}\in\tAID$)
a message type $\dval{mt}\in\tMT$ and a number $\dval{k}\in\NN$ is stored, indicating the
message type of the message received last and number of fragments received so far.
Additionally, the function stores the concatenated data $\dval{d}\in \tDATA$ from these fragments.
In the formal model (and indeed in any implementation) we need to
extract information form \queues. To this end, we define the 
functions\footnote{In contrast to \idtab, the function \queues\ will change during protocol execution. That is why we add it as argument to 
the extractions.}
$\fnmtype:\tQUEUES\times \tAID \to  \tMT$, 
$\fnframeid:\tQUEUES\times \tAID\to\NN$, and
$\fncontents:\tQUEUES\times \tAID \to  \tDATA$
 by
\[\begin{array}{r@{\hspace{0.5em}}c@{\hspace{0.5em}}l}
\mtype{\dval{store}}{\dval{aid}} &:=& \pi_{1}(\dval{store}(\dval{aid}))\ ,\\ 
\frameid{\dval{store}}{\dval{aid}} &:=& \pi_{2}(\dval{store}(\dval{aid}))\ ,\\ 
\contents{\dval{store}}{\dval{aid}} &:=& \pi_{3}(\dval{store}(\dval{aid}))\ .\\ 
	     \end{array}\]
The function $\fnmtype$ returns the message type handled at the moment for sender $\dval{aid}$;
$\fnframeid$ extracts the number of fragments a node has received so far; and 
the data received so far is accessed by $\fncontents$.

The main intention of the reassembly protocol is to  reassemble data received via the {\can}. 
For this purpose we define the function 
\[
\fnappenddata:\tDATA\times\tDATA\to\tDATA\ ,
\]
which concatenate two strings of data.%
\footnote{The index $8$ is only a reminder that we append $8$ bytes to a data string, and 
to indicate that this function follows the same lines as the functions $\fnheaddata$ and 
$\fntaildata$.}

The following table summarises the entire data structure we use for the reassembly protocol.

{\centering{\small
\setlength{\tabcolsep}{4.0pt}
\begin{longtable}{@{}|l|l|l|@{}}
\hline
\textbf{Basic Type} & \textbf{Variables} & \textbf{Description}\\
\hline
 \tMSG		&\msg	&messages\\
 \tDATA		&\data	&stored data\\
 \NN		&\no	&fragment counter\\
 \tMT		&\mt	&message types\\
 \tCID 		&	& CAN IDs\\
 \tAID		&\sip	&unique sender (application) identifier\\
\hline
\hline
\textbf{Complex Type} & \textbf{Variables} & \textbf{Description}\\
\hline
$\tQUEUES \triangleq \tAID \to $ &\queues		&storage of received fragments\\
$\hspace{1.02cm} \tMT \times \NN \times\; \tDATA$ &&\\
\hline
\hline
\multicolumn{2}{|l|}{\textbf{Constant/Predicate}}& \textbf{Description}\\
\hline
\multicolumn{2}{|l|}{$\emptystring:\tDATA$}&
	the empty data string\\
\multicolumn{2}{|l|}{$\mtundef:\tMT$}&
	special message type symbol, denoting undefined message type\\
\multicolumn{2}{|l|}{$\R: \tSID$}& reassembling protocol identifier for hardware component $H$\\
\hline
\hline
\multicolumn{2}{|l|}{\textbf{Function}} & \textbf{Description}\\
\hline
\multicolumn{2}{|l|}{$\canmID : \tCID\times\tDATA \to \tMSG$}&
	create CAN messages out of identifier and payload\\
\multicolumn{2}{|l|}{$\fncanid: \tMT\times\NN \rightharpoonup \tCID$}&
	returns a CAN ID for a given message type and a counter\\
\multicolumn{2}{|l|}{$\fnsender:\tCID\to\tAID$}&
	determines unique sender ID for a particular message type\\	
\multicolumn{2}{|l|}{$\fnmtype:\tQUEUES\times \tAID \to  \tMT$}&
	the message type of the message currently reassembled \\
\multicolumn{2}{|l|}{$\fnframeid:\tQUEUES\times \tAID\to\NN$}&
	indicates the number of fragments received so far\\
\multicolumn{2}{|l|}{$\fncontents:\tQUEUES\times \tAID \to  \tDATA$}&
	returns the (reassembled) data received so far\\
\multicolumn{2}{|l|}{$\fnsize : \tCID\to \NN$}&
	number of CAN messages into which a message is split\\
\multicolumn{2}{|l|}{$\fndatamsg : \tMT\times\tDATA\to\tDATA$}&
	create data packet to be delivered to application\\
\multicolumn{2}{|l|}{$\fnappenddata:\tDATA\times\tDATA\to\tDATA$}&
	concatenates data strings\\
\hline
\caption[]{\rule[10pt]{0pt}{3pt}Data structure for the Reassembly Protocol}
\end{longtable}
}}

\vspace{-5mm}
\subsubsection{Formal Specification}\label{sec:reass_spec}

The process $\RASS$ models all events that  occur
after a CAN message is received by a node. This includes
the reassembly of messages as well as their delivery.             
This process maintains a data variable {\queues},  in which it stores
received data fragments that await further action.

  \algsetup{linenodelimiter=.,linenosize=\tiny}
  \begin{algorithm}[ht]
    {\footnotesize
      \caption{Reassembly}
      \label{pro:rass}
      \begin{algorithmic}[1]
        % !TEX root = ../paper.tex
\DEFPROCESS{\RASS}{\queues}																															\label{rass:line0}
    \receiveL{\msg}\ .                                                                                                                                					\label{rass:line1}          
    \IF[distill {\mt}, {\no} and {\data} from CAN fragment]{$\msg = \canm{\canid{\mt}{\no}}{\data} $}                       		\label{rass:line2}
	\UPD{\sip:=\sender{\mt}}																																	\label{rass:line3}
        \PAR																																								\label{rass:line4}
        \IF[new message]{$\no\mathbin=1$}																												\label{rass:line5}
		\PAR																																								\label{rass:line6}
		\IF[full message received (consists of one fragment only)]{$\size{\mt}\mathbin=1$}										\label{rass:line7}
			\deliverL{\datamsg{\mt}{\data}}\ .																												\label{rass:line8}
			\UPD{\queue{\sip}:=(\mtundef,0,\emptystring)} \COM{erase data from the data storage}								\label{rass:line9}
			\rassL{\queues}																																			\label{rass:line10}
		\ELSIF[fragment to be stored]{$\size{\mt}\mathbin>1$}																					\label{rass:line11}
			\UPD{\queue{\sip}:=(\mt,1,\data)}																								\label{rass:line12}
			\rass{\queues} 																					
		\ENDIFii
		\ENDPAR																																						\label{rass:line13}
	\ELSIF{$\no\mathbin{\not=}1
	                   \wedge\mt=\mtype{\queues}{\sip}\wedge\no\mathbin=\frameid{\queues}{\sip}{+}1$}                         	\label{rass:line14}   
		\COMLINE{message needs reassembly}																										\label{rass:line15}
		\UPD{\queue{\sip}:=(\mt,\no,\appenddata{\contents{\queues}{\sip}}{\data})}            											\label{rass:line16}
                 \PAR																						\label{rass:line17}
                 \IF[full message received]{$\frameid{\queues}{\sip}\mathbin=\size{\mt}$}                               					\label{rass:line18}
                 	\deliverL{\datamsg{\mt}{\contents{\queues}{\sip}}}\ .																			\label{rass:line19}
			\UPD{\queue{\sip}:=(\mtundef,0,\emptystring)}\COM{erase data from the data storage}								\label{rass:line20}
			\rassL{\queues}																																			\label{rass:line21}
		\ELSIF[message not yet complete]{$\frameid{\queues}{\sip}\mathbin{\not=}\size{\mt}$}									\label{rass:line22}
			\rassL{\queues}																																			\label{rass:line23}
		\ENDIFii
		\ENDPAR																																						\label{rass:line24}
	\ELSIF[repeated fragment]{$\no\not=1\ \wedge\ 
											(\mt\mathbin=\mtype{\queues}{\sip}\wedge \no\mathbin=\frameid{\queues}{\sip})$}   	\label{rass:line25}
		\rassL{\queues} \COM{ignore repeated fragment}																							\label{rass:line26}

	\ELSIF{$\no{\not=}1\mathbin\wedge(\mt{\not=}\mtype{\queues}{\sip}\mathbin\vee
                       (\no{\not=}\frameid{\queues}{\sip}{+}1\mathbin\wedge\no{\not=}\frameid{\queues}{\sip}))\!$} 				\label{rass:line27}
		\COMLINE{non-initial  fragment out of order}																									\label{rass:line28}
		\UPD{\queue{\sip}:=(\mtundef,0,\emptystring)}																								\label{rass:line29}
		\rassL{\queues}                                                 																								\label{rass:line30}

	\ENDIFii
        \ENDPAR																																						\label{rass:line31}
    \ENDIFii

	\end{algorithmic}
    }%end{footnotesize}
  \end{algorithm}

First, the message has to receive a CAN messages from the CAN driver (\receive{\msg}).
After that, the process
$\RASS$ extracts from it the $\data$ received;
it also distils the message type $\mt$ and
fragment counter $\no$ from the CAN ID (Line~\ref{rass:line2}).
From the message type the process also determines the sender $\sip$ of the 
message (Line~\ref{rass:line3}).
As long as a node did not receive all fragments of a message, it cannot deliver 
the (reassembled) message. Hence upon arrival fragments are sorted by their 
sender and the
carried data is concatenated to any existing data belonging to the same message received
before. There need to be as many data queues as there are senders,
or as many as there are senders that can potentially send messages to the current node.
As soon as all fragments of a fragmented message are received, the message can be passed on to the application layer. 

Lines~\ref{rass:line5}--\ref{rass:line13} cover the case when a first fragment of a split message arrives; 
here it is sufficient to check the fragment number \no.
The process then checks whether the CAN message under consideration is just an ordinary CAN message or a fragment (which needs reassembling). 
In the former case (Lines~\ref{rass:line7}--\ref{rass:line10})%
\footnote{As stated earlier a $1$ in the {\fnsize} field of the identifier table indicates a standard CAN message (for example a legacy message).}
the message is delivered to the application layer
(Line~\ref{rass:line8}), the local storage is erased (Line~\ref{rass:line9}) and the process returns to its main routine (Line~\ref{rass:line10}). 
If the message received is the first fragment of a longer (split) message, the process just stores the fragment (Line~\ref{rass:line12}).

In case the fragment received is not the first fragment, the process 
checks whether the received fragment fits in the sequence of received messages. This is achieved 
by the fragment counter \no, using $\no\mathbin=\frameid{\queues}{\sip}{+}1$.
Since we assume that messages are not reordered (see \autoref{sec:assum}), this counter must have a value that is exactly one
higher than the one of the previous message.
Line~\ref{rass:line14} also checks the message type. If the message type does not fit the stored one, 
the fragment, although it might have the correct fragment number, does not fit.
If these checks evaluate to true (Line~\ref{rass:line14}) the received
data is concatenated to the data received before from the same sender; 
and the fragment counter is incremented, which is done by storing {\no} as second component (Line~\ref{rass:line16}).
After this update, the process checks whether all fragments of a split message have been received. 
If this is the case ($\frameid{\queues}{\sip}=\size{\mt}$) the message is delivered and the data is erased from the data storage. 
In case there are still fragments missing, the process returns to the main routine, awaiting new messages. 

The remaining lines (Lines~\ref{rass:line25}--\ref{rass:line30}) handle the case 
where an out-of-order fragment is received.
If a fragment is received with the same message type and the same fragment counter ($\not=1$)
as the last fragment, 
the protocol assumes that it is a repeated fragment and ignores the entire message.
If fragments of a message were lost or reordered, the reassembly protocol will fail---a check 
for this is implemented in Line~\ref{rass:line27}.
The data received so far is erased and the entire message will be ignored.

Note that when an application cancels a message before the fragmentation protocol has generated
all fragments, only a prefix of the full sequence of fragments will ever reach the reassembly
protocol. In that case the incomplete message will sit in the slot allocated to its sender until a
first fragment of a new message from the same sender arrives. At that time Line~\ref{rass:line5}
of Process~\ref{pro:rass} will be invoked, and the incomplete message is discarded
(Line~\ref{rass:line9} or~\ref{rass:line12}).

\subsubsection{Initialisation}\label{ssec:reass_init}
The reassembly protocol $R_H$ is initialised by 
$\R\mathop{:}(\xi,\rass{\queues})$, 
where $\R$ is the identifier of the local instance of the reassembly protocol, and
$\xi(\queues(\sip))=(\mtundef,0,\emptystring)$
for all applications \mbox{$\sip\in\tAID$}.

\subsection{The Multiplexer}\label{sec:multiplexer}
The multiplexer combines several instances of the fragmentation protocol with the CAN driver and the CAN controller. In particular it can receive messages from multiple fragmentation protocols and uses an internal prioritisation mechanism to forward the most important messages. 
It also checks whether some CAN messages should be erased from the TX buffers (out-queue of the CAN controller) as soon as a new CAN message arrives. 
By this mechanism we avoid the example where high-priority messages are blocked by low-priority messages (cf.\ \autoref{sec:can}). 

\begin{figure}
\centering
\includegraphics[width=7cm]{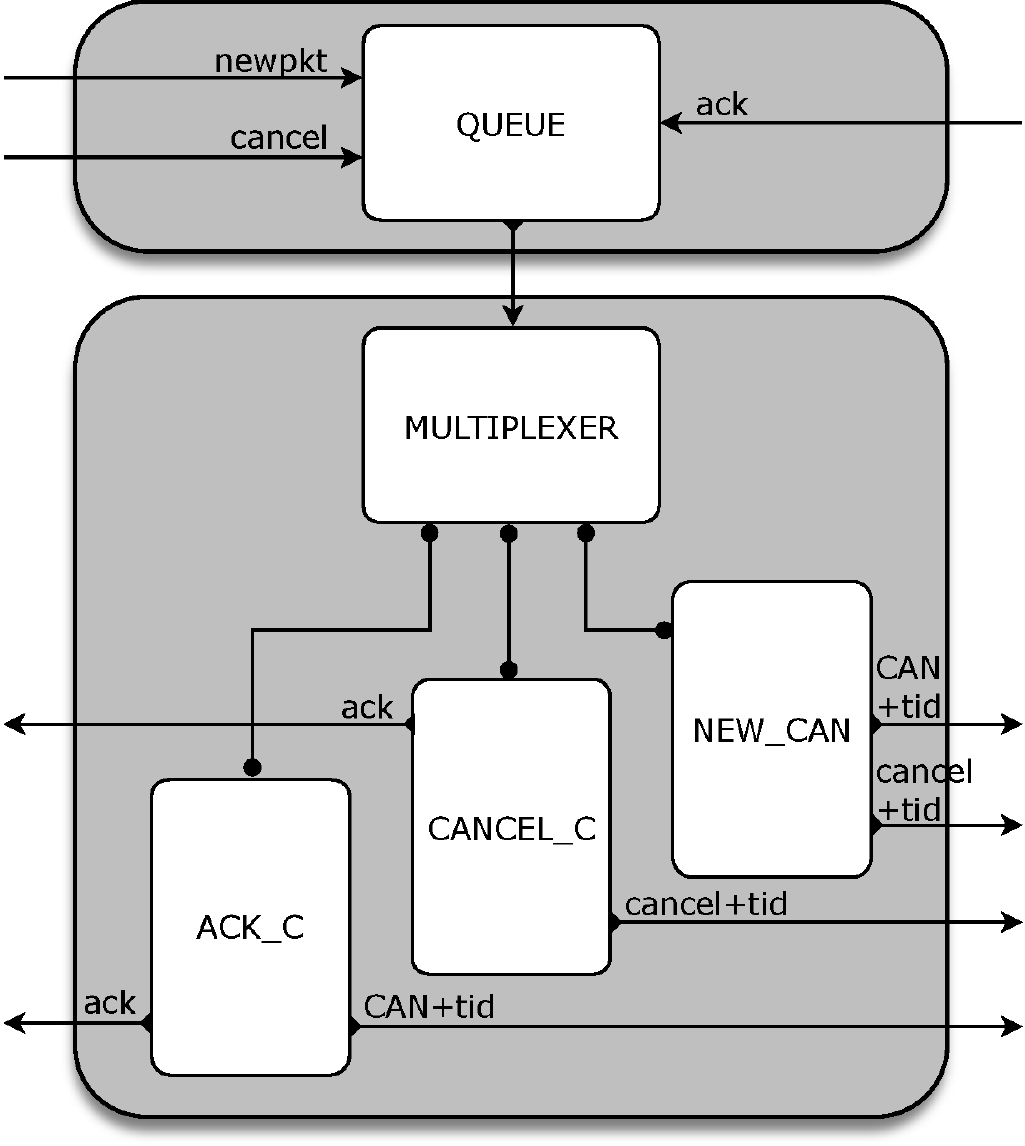}
\caption{Structure of the Multiplexer}
\label{fig:message_mult}
\end{figure}

The protocol consists of $5$ processes: \MULT, \TREQ, \CCANC, \ACKC\ and \QMSG.
\begin{itemize}
\item The process \MULT\ is similar to \FRAG. It is the basic process that receives messages, and, depending on the type of the message, calls other processes. 
When there is no message handling going on, it idles until a new message arrives.
\item The process \TREQ\ handles a new CAN message received from a fragmentation protocol.
In particular it stores the CAN message and determines whether the message 
is important enough to be forwarded straight to the CAN driver. 
\item The process \CCANC\ handles all actions to be performed when a cancellation message arrives from the fragmentation protocol.
\item The process \ACKC\ reacts on incoming acknowledgements, stemming from the CAN driver. 
It forwards the acknowledgement to the fragmentation protocol to inform it that a new fragment could be sent; 
at the same time it sends another CAN message to the CAN driver, in case there is one in the local storage.
\item  The last process $\QMSG$ concerns message handling. Whenever a message is received,
	  it is first stored in a message queue. If the multiplexer is able to handle a message
	  it pops the oldest message from the queue and handles it. 
This process ensures that the multiplexer is input enabled, as required in \autoref{sec:can_spec}.
\end{itemize}

\subsubsection{Data Structure}\label{sec:multiplexer_data2}
The multiplexer has to store all incoming CAN messages, which were generated by instances of the fragmentation protocol.
Since we assume that each CAN ID determines a unique sender, and that each application is sending only one message a time, 
it suffices to implement an array that maps CAN IDs to messages. 
\[\begin{array}{r@{\hspace{0.5em}}c@{\hspace{0.5em}}l}
		    	  \PQUEUE \triangleq \tCID&\to& \tMSG
	     \end{array}\]
Our formal specification leaves the implementation details open and defines
$\pqueue\in\PQUEUE$ as a function. An implementation can be a function, an array, or a priority queue---it is 
the latter that has been implemented in our research vehicle.

The multiplexer also keeps track of the messages currently stored in the TX buffers.
We assume a set $\tTX$ of TX-buffer identifiers; the variables $\txid$ and $\wtx$ range over $\tTX$.
It is sufficient to store the (unique) CAN ID together with a flag that indicates
whether the CAN controller is currently cancelling the
message of the indicated TX buffer. To this end we define a function of type $\TXMIRROR$, where
\[\begin{array}{r@{\hspace{0.5em}}c@{\hspace{0.5em}}l}
		    	  \TXMIRROR \triangleq \tTX&\to& \tCID\times \BB\ .
	     \end{array}\]
For a function $\dval{txs}\in\TXMIRROR$ we define projections to access the first and second components. 
\[\begin{array}{r@{\hspace{0.5em}}c@{\hspace{0.5em}}l}
			\fntxcid: \tTX\times\TXMIRROR &\to& \tCID\\
		    	  \txcid{\dval{tid}}{\dval{txs}} & := & \pi_{1}(\dval{txs}(\dval{tid}))\\[2mm]
			\fntxab: \tTX\times\TXMIRROR &\to& \BB\\
		    	  \txab{\dval{tid}}{\dval{txs}} & := & \pi_{2}(\dval{txs}(\dval{tid}))
	     \end{array}\]
The first function distils the CAN ID of a message that is currently stored in the TX buffer with identifier $\dval{tid}$;
the second one signals whether a cancellation message has been sent to the corresponding  TX buffer.
Moreover, we define a function that collects all CAN IDs that are currently stored in the TX buffers.
\[\begin{array}{r@{\hspace{0.5em}}c@{\hspace{0.5em}}l}
			\fncansintxs: \TXMIRROR &\to& \pow(\tCID)\\
		    	  \cansintxs{\dval{txs}} & := &
                          \{\dval{cid}\mid\exists \dval{tid} :
                          \txcid{\dval{tid}}{\dval{txs}} = \dval{cid} \neq \cidundef\}
	     \end{array}\]

Finally we will use a partial function $\fnworstmbid : \TXMIRROR\rightharpoonup \tTX$ 
that determines the name of the TX buffer with the least urgent message (the one with the largest CAN ID):
\[
\worstmbid{\dval{txs}} = \dval{tid}
\ \Leftrightarrow\
\txcid{\dval{tid}}{\dval{txs}} = \max(\cansintxs{\dval{txs}})
	\ .\]

When a TX buffer becomes available, we want to select a new message from the priority queue to send
next---the one with the smallest CAN ID that is not already forwarded to the TX buffers:
\vspace{-.5mm}
\[
\begin{array}{r@{\hspace{0.5em}}c@{\hspace{0.5em}}l}
			\fnnewtask: \PQUEUE\times \TXMIRROR&\to& \tCID\\
		    	 \newtask{\dval{prio},\dval{txs}} & := & \min\{\dval{cid}\mid
                         \dval{prio}(\dval{cid})\not=\cidundef \wedge \dval{cid}\notin
                         \cansintxs{\dval{txs}}\}
\end{array}\]
where $\min\{\}$ is defined to be the special element $\cidundef$.  By this latter definition
$\fnnewtask$ becomes a total function; $\newtask{\dval{prio},\dval{txs}} =\cidundef$
indicates that no message needs to be scheduled next.

To implement a proper prioritisation mechanism and to avoid the blocking example of \autoref{sec:can}  
a multiplexer has to determine the $n$ CAN messages in the priority queue with highest priority (lowest CAN identifier)---these 
messages should be sent as soon as possible. Here $n$ corresponds to the number $\noTX$ of available TX buffers.

In case there is only one TX buffer, as it is the case for the mission board, 
we can define a partial function similar to $\fnbest$ (\autoref{sec:can_data}) that determines the identifier of the 
CAN message with highest priority currently stored in $\dval{prio}\in\PQUEUE$:
\[\begin{array}{cl@{}l}
&\multicolumn{2}{l}{\nbesti{\dval{prio}} = \dval{cid}}\\
\Leftrightarrow&
\exists \dval{d}: \big(&
	\dval{prio}(\dval{cid}) = \canm{\dval{cid}}{\dval{d}}\ \wedge \\
	&&  (\exists \dval{cid}', \dval{d}': \dval{prio}(\dval{cid}') = \canm{\dval{cid}'}{\dval{d}'} \Rightarrow \dval{cid}\leq \dval{cid}')\big)\ .
\end{array}\]

In case of $n$ TX buffers we require a function that determines $n$ messages. The easiest way
is to define a recursive function. The base case ($n=1$) is a total function corresponding to the function $\fnnbesti$:
\[\begin{array}{r@{\hspace{0.5em}}c@{\hspace{0.5em}}l}
		    	  \fnnbest : \NN\times\PQUEUE&\to& \pow(\tCID)\\
			   \nbest[1]{\dval{prio}} &:=&\left\{
			   		\begin{array}{ll}
						\{\nbesti{\dval{prio}}\} &\mbox{if $\nbesti{\dval{prio}}$ is defined}\\
						\{\}&\mbox{otherwise\ .}
					\end{array}
			   \right.
	     \end{array}\]
Here, the second case describes the situation of an empty queue \pqueue.

\noindent The recursive case ($n>1$) is then defined as 
\[
	\nbest[n]{\dval{prio}} = \nbest[1]{\dval{prio}} \cup \nbest[n\mathop-1]{\dval{prio}'}\ ,
\]
where $\dval{prio}'(\dval{cid}) = \dval{prio}(\dval{cid})$ for all $\dval{cid}\not=\nbesti{\dval{prio}}$ and 
$\dval{prio}'(\nbesti{\dval{prio}}) = \msgundef$ (in the case that $\nbesti{\dval{prio}}$ is defined).

This recursive definition seems to be overly complicated. In fact when implementing a priority queue a simple position argument 
is enough to check whether a given CAN message $\dval{msg}$ is among the $n$ messages with highest priority: just insert 
$\dval{msg}$ into the queue and check if its position is smaller than $n$. Since we want to abstract from implementation details,
we chose, however, to present a recursive definition which allows different implementations. 

In the specification presented below we have to determine whether a new message should be moved to a TX buffer immediately. 
Since the number $\noTX$ of TX buffers is a constant, we define a function 
\[\begin{array}{r@{\hspace{0.5em}}c@{\hspace{0.5em}}l}
		    	  \fnnbest : \PQUEUE&\to& \pow(\tCID)\\
			   \nbest{\dval{prio}} &:=& \nbest[\noTX]{\dval{prio}}\ .
	     \end{array}\]
As for the CAN receiver, we use 
a queue-style data structure for modelling an inbox queue. 
In particular we make again use of the
standard functions
$\fnhead:[\tMSG]\rightharpoonup\tMSG$,
$\fntail:[\tMSG]\rightharpoonup[\tMSG]$ and
$\fnappend:\tMSG\times[\tMSG]\rightarrow[\tMSG]$.	  
	     
The following table summarises the entire data structure we use for the multiplexer.
{\centering{\small
\setlength{\tabcolsep}{4.0pt}
\begin{longtable}{@{}|l|l|l|@{}}
\hline
\textbf{Basic Type} & \textbf{Variables} & \textbf{Description}\\
\hline
 \tMSG		&\msg	&messages\\
 \tDATA		&\data	&stored data\\
 \tTX		&\txid, \wtx	&identifiers of TX buffers\\
 \BB 		&\abort, \ackb 	&Boolean flags\\
 \tCID		&\cid, \bid 	&CAN IDs\\
 \tSID		&	&CAN software component identifiers\\ 
\hline
\hline
\textbf{Complex Type} & \textbf{Variables} & \textbf{Description}\\
${[\tMSG]}$						&\msgs			&message queues\\
$\PQUEUE \triangleq \tCID\to \tMSG$&\pqueue	&	priority queue for CAN messages\\
 $\TXMIRROR \triangleq \tTX\to \tCID\times \BB$&\txmirror& array of CAN IDs currently in TX buffer\\
\hline
\hline
\multicolumn{2}{|l|}{\textbf{Constant/Predicate}}& \textbf{Description}\\
\hline
\multicolumn{2}{|l|}{$\msgundef:\tMSG$}&special message symbol (indicating absence of a message)\\
\multicolumn{2}{|l|}{$\cidundef:\tCID$}&
	special CAN ID symbol, denoting undefined CAN ID\\
\multicolumn{2}{|l|}{${[\,]}:{[\tMSG]}$}&
	empty queue\\
\multicolumn{2}{|l|}{$\C: \tSID$}& sending CAN driver identifier for hardware component $H$\\
\hline
\hline
\multicolumn{2}{|l|}{\textbf{Function}} & \textbf{Description}\\
\hline
\multicolumn{2}{|l|}{$\canceliID : \tCID \to \tMSG$} & 
	cancellation message from fragmentation protocol\\
\multicolumn{2}{|l|}{$\cancelID : \tMSG$} & 
	cancellation message to CAN driver\\
\multicolumn{2}{|l|}{$\ackID : \BB\to \tMSG$}&
	acknowledgement (from CAN driver and to fragm.\ prot.) \\
\multicolumn{2}{|l|}{$\canmID : \tCID\times\tDATA \to \tMSG$}&
	create CAN messages out of identifier and payload\\
\multicolumn{2}{|l|}{$\fnamsg:\tTX\times\tMSG \to \tMSG$}&
  wrapper function to add a TX-identifier to a message\\
\multicolumn{2}{|l|}{$\fnnewtask: \PQUEUE \to \tCID$}&
  selecting the most urgent CAN ID from priority queue\\
\multicolumn{2}{|l|}{$\fnnbest : \PQUEUE\to \pow(\tCID)$}&
  returns CAN IDs that should be scheduled\\
\multicolumn{2}{|l|}{$\fntxcid : \tTX\times\TXMIRROR\rightharpoonup \tCID$}&
  distils CAN ID of message in TX buffer\\
\multicolumn{2}{|l|}{$\fntxab : \tTX\times\TXMIRROR\rightharpoonup \BB$}&
  signals whether cancellation signal was sent to TX buffer\\
\multicolumn{2}{|l|}{$\fnworstmbid : \TXMIRROR\rightharpoonup \tTX$}&
  returns TX buffer with least urgent message\\
\multicolumn{2}{|l|}{$\fnhead:[\tMSG]\rightharpoonup\tMSG$}&
	returns the `oldest' element in the queue\\
\multicolumn{2}{|l|}{$\fntail:[\tMSG]\rightharpoonup[\tMSG]$}&
	removes the `oldest' element in the queue\\
\multicolumn{2}{|l|}{$\fnappend:\tMSG\times[\tMSG]\rightarrow[\tMSG]$}&
	inserts a new element into the queue\\
\multicolumn{2}{|l|}{$\fnfragname:\tCID \to \tSID$}&
  gives name of fragmentation protocol handling a message\\
\hline
\caption[]{\rule[10pt]{0pt}{3pt}Data structure for the Multiplexer}
\end{longtable}
}}

\subsubsection{Formal Specification}\label{sec:multiplexer_spec}
\myparagraph{The Main Loop}
The basic process \MULT\ (\Pro{multiplexer}) receives
messages from the fragmentation protocol or the CAN driver.
Since the multiplexer is not always ready to receive messages, we equip the process with an in-queue (see below);
so technically the multiplexer receives a message from this queue.
\MULT\ maintains two data variables {\pqueue} and {\txmirror}.
The former implements a priority queue which contains all CAN messages to be sent via 
the {\can};
the later is a local storage which keeps track of the CAN IDs
currently sent by or stored in the 
TX buffers of the CAN controller. 

  \algsetup{linenodelimiter=.,linenosize=\tiny}
  \begin{algorithm}[ht]
    {\footnotesize
      \caption{Multiplexer---Main Loop}
      \label{pro:multiplexer}
      \begin{algorithmic}[1]
        % !TEX root = ../paper.tex
\DEFPROCESS{\MULT}{\pqueue\comma\txmirror}													\multilabel{line0} 
	\receiveL{\msg}\ .                         																			\multilabel{line1}                                                   
	\PAR                         																								\multilabel{line2}      
	\IF[new fragment]{$\msg = \canm{\cid}{\data}$}     													\multilabel{line3}     
		\treqL{\msg}{\pqueue}{\txmirror}																			\multilabel{line4}      
	\ELSIF[cancellation message received]{$\msg = \canceli{\cid}$} 							\multilabel{line5}      
		\ccancL{\cid}{\pqueue}{\txmirror}																		\multilabel{line6}      
	\ELSIF[message from CAN controller]{$\msg = \amsg{\txid}{\ackM{\ackb}}$}			\multilabel{line7}      
		\ackcL{\ackb}{\txid}{\pqueue}{\txmirror}																\multilabel{line8}      
	\ENDIFii
	\ENDPAR																												\multilabel{line9}      

	\end{algorithmic}
    }%end{footnotesize}
  \end{algorithm}

First, as usual, a message has to be received (Line~\ref{appmulti:line1}). 
After that, the process \MULT\ checks the type of the message and calls a process that can handle this message: 
in case a CAN message is received from the fragmentation protocol, the process
\TREQ\ is called (Line~\ref{appmulti:line4});  
in case of an incoming cancellation request the process \CCANC\ is executed (Line~\ref{appmulti:line6});
and in case a message from the CAN driver is read, the process \ACKC\ is called (Line~\ref{appmulti:line8}).
%\outP{carrying the Boolean flag {\ackb} and the name of the TX buffer ($\txid$) received from the CAN driver.}
In case a message of any other type is received, the process
\MULT\ deadlocks; it is a proof obligation to check that this will not occur.

\myparagraph{New CAN Message}
In case a new CAN message is sent from an instance of the fragmentation protocol, 
the process {\TREQ} stores the CAN message and determines whether the newly received message 
is important enough to be forwarded directly to the CAN driver. The formal specification is 
shown in \Pro{newcan}.

  \algsetup{linenodelimiter=.,linenosize=\tiny}
  \begin{algorithm}[ht]
    {\footnotesize
      \caption{New CAN Message Received}
      \label{pro:newcan}
      \begin{algorithmic}[1]
        % !TEX root = ../paper.tex

\DEFPROCESS{\TREQ}{\msg\comma\pqueue\comma\txmirror}																			\treqlabel{line0}
	\IF[distill $\cid$ out of $\msg$]{$\msg = \canm{\cid}{\data}$}
	\UPD{\pqueueupd{\cid} := \msg} 	\COM{store message in priority queue}															\treqlabel{line1}
	\PAR
	\IF[message should be scheduled]{$\cid\in\nbest{\pqueue}$}																				\treqlabel{line2}
		\PAR																																								\treqlabel{line3}
		\IF[TX buffer $\txid$ is free]{$\txcid{\txid}{\txmirror} = \cidundef$}																	\treqlabel{line4}	
			\UPD{\mboxupd{\txid}:=(\cid,\false)}  																											\treqlabel{line5}
			\unicastL{\C}{\amsg{\txid}{\msg}}\ .\  \COM{pass message to CAN	driver, to put in free slot}						\treqlabel{line6}
			\multiL{\pqueue}{\txmirror}																															\treqlabel{line7}
		\ELSIF[cancel message with lowest priority]{$\forall \txid\in\tTX: \txcid{\txid}{\txmirror} \not= \cidundef$}			\treqlabel{line8}
			\PAR																																							\treqlabel{line9}
				\UPD{\wtx := \worstmbid{\txmirror}}	\COM{identify TX buffer containing lowest CAN ID}						\treqlabel{line10}
				\PAR
				\IF[TX buffer $\wtx$ is still active]{$\txab{\wtx}{\txmirror} = \false$}															\treqlabel{line11}
                    \UPD{\mboxupd{\wtx}:=(\txcid{\wtx}{\txmirror},\true)}  \COM{set the abort-flag of buffer $\wtx$}            \treqlabel{line12}
					\unicastL{\C}{\amsg{\wtx}{\cancel}}\ .\		\COM{cancel contents of buffer $\wtx$}								\treqlabel{line13}
					\multiL{\pqueue}{\txmirror}																													\treqlabel{line14}
				\ELSIF[TX was already asked to clean up]{$\txab{\wtx}{\txmirror} = \true$}												\treqlabel{line15}
					\multiL{\pqueue}{\txmirror}																													\treqlabel{line16}
				\ENDIFii
				\ENDPAR						
			\ENDPAR																																					\treqlabel{line17}
		\ENDIFii
		\ENDPAR																																						\treqlabel{line18}
	\ELSIF[message not important enough to be scheduled right now]{$\cid\not\in\nbest{\pqueue}$}								\treqlabel{line19}
		\multiL{\pqueue}{\txmirror}																																\treqlabel{line20}
    \ENDIFii
    		\ENDPAR
    \ENDIFii

	\end{algorithmic}
    }%end{footnotesize}
  \end{algorithm}

The received CAN message is first stored in the queue {\pqueue} (Line~\ref{apptreq:line1}), which contains all messages to be sent via the 
{\can}. The protocol just stores the newly received message, it does not check for emptiness 
of $\pqueueupd{\cid}$. Therefore, to guarantee that no message is lost 
the property $\pqueueupd{\cid}=\msgundef$ needs to hold before Line~\ref{apptreq:line1} is executed;
it needs to be proven.
The protocol then determines whether the message should directly be forwarded to the CAN driver---this is the case 
if the CAN ID is among the $n$ messages with lowest CAN IDs currently stored in {\pqueue} (Line~\ref{apptreq:line2}). Here $n$ equals 
the number $\noTX$ of TX buffers available in the CAN controller. 
Lines~\ref{apptreq:line3}--\ref{apptreq:line18} present all actions to be performed in case the message 
is forwarded to the CAN driver. 

In case there exists an empty TX buffer $\txid$, which is currently not used,
the message should be sent to this TX buffer, and there is no need to erase a used TX buffer. 
The empty buffer $\txid$ is chosen in Line~\ref{apptreq:line4}.\footnote{Since $\txid$ is a free variable, 
it will be instantiated with a value that validates
$\txcid{\txid}{\txmirror} = \cidundef$;
so the condition in the guard is satisfied iff $\exists \txid\in\tTX: \txcid{\txid}{\txmirror} = \cidundef$.}
The CAN message is then forwarded to the connected CAN driver {\C} in Line~\ref{apptreq:line6}. Since the CAN driver needs also the 
name of the TX buffer to be used, the value $\txid$ is sent next to the CAN message $\msg$.
The multiplexer also updates the local variable $\txmirror$ (Line~\ref{apptreq:line5}), which keeps track 
of those CAN identifiers that are currently sent by or stored in the TX buffers. By this, the newly received message 
has been handled and the process can return to the main routine (Line~\ref{apptreq:line7}).

In case all available TX buffers are used (Line~\ref{apptreq:line8}), the least important message---the CAN message with the largest
CAN ID---needs to be removed from the TX buffer and rescheduled later. 
This avoids the blocking example presented earlier.
In Line~\ref{apptreq:line10} the process {\TREQ} determines the name of the TX buffer that contains the `worst' message currently 
handled for sending. The CAN message that should be stored in this
particular TX buffer cannot be put there immediately; a cancellation request needs to be sent first, 
and an acknowledgement needs to be received that informs the multiplexer about a free TX buffer.
The routine checks whether a cancellation request was sent earlier, using the 
function $\fntxab$. If this is the case, it returns straight to the process \MULT; otherwise 
a cancellation message is sent to the CAN driver \C, identifying the TX buffer that needs cancellation (Line~\ref{apptreq:line13}).

If the newly received CAN message is not important enough to be forwarded to the CAN driver immediately 
(Line~\ref{apptreq:line19}), the process {\TREQ} just returns to the main process (Line~\ref{apptreq:line20}), where it
awaits a new message. The stored message will be handled later when a TX buffer becomes available.

\myparagraph{Cancellation Message} 
All actions to be taken if a cancellation request is received are formalised in \Pro{ccanc}.
Any cancellation message received carries the CAN ID {\cid} of the message to be cancelled.

  \algsetup{linenodelimiter=.,linenosize=\tiny}
  \begin{algorithm}[ht]
    {\footnotesize
      \caption{Cancellation Message Received}
      \label{pro:ccanc}
      \begin{algorithmic}[1]
        % !TEX root = ../paper.tex
\DEFPROCESS{\CCANC}{\cid\comma\pqueue\comma\txmirror}											\label{ccanc:line0}
	\IF[nothing to cancel]{$\pqueueupd{\cid}=\msgundef$}														\label{ccanc:line1}
		\multiL{\pqueue}{\txmirror}																								\label{ccanc:line2}
	\ELSIF[send cancellation message to CAN driver]{$\pqueueupd{\cid}\not=\msgundef$}				\label{ccanc:line3}
		\UPD{\pqueueupd{\cid}:=\msgundef}																				\label{ccanc:line4}
		\PAR																																\label{ccanc:line5}
		\IF[determine TX buffer]{$\txcid{\txid}{\txmirror}=\cid$}													\label{ccanc:line6}
%		\COMLINE{\chP{an alternative is $\cid\in\cansintxs{\txmirror}$}}
			\PAR																															\label{ccanc:line7}
			\IF[cancellation already sent]{$\txab{\txid}{\txmirror} = \true$}										\label{ccanc:line8}
				\multiL{\pqueue}{\txmirror}																						\label{ccanc:line9}	
			\ELSIF{$\txab{\txid}{\txmirror} = \false$}																		\label{ccanc:line10}
%				\UPD{\outP{\txab{\txid}{\txmirror} := \true}}																	\label{ccanc:line11}
				\UPD{\mboxupd{\txid}:=(\txcid{\txid}{\txmirror},\true)}																\label{ccanc:line11}
				\unicastL{\C}{\amsg{\txid}{\cancel}}\ .\				 													\label{ccanc:line12}
				\multiL{\pqueue}{\txmirror}																						\label{ccanc:line13}
			\ENDIFii
			\ENDPAR																													\label{ccanc:line14}
		\ELSIF[message not in TX]{$\forall \txid\in\tTX: \txcid{\txid}{\txmirror}\not=\cid$}				\label{ccanc:line15}
%		\COMLINE{\chP{an alternative $\cid\not\in\cansintxs{\txmirror}$}}
			\unicastL{\fragname{\cid}}{\ackM{\false}}\ .\ 																\label{ccanc:line16}
			\multiL{\pqueue}{\txmirror}																							\label{ccanc:line17}
		\ENDIFii
	\ENDPAR																															\label{ccanc:line}
	\ENDIFii

	\end{algorithmic}
    }%end{footnotesize}
  \end{algorithm}

In case the multiplexer has previously handled the message
already, the value of $\pqueueupd{\cid}$ is $\msgundef$
(Line~\ref{ccanc:line1}). This situation can happen if for example the message was sent successfully, but the 
acknowledgement was not received by the time the
fragmentation protocol requested the cancellation.
In this case, the process has nothing to do, ignores the message and returns to the main process \MULT\ (Line~\ref{ccanc:line2}).

In case the message that needs cancellation is still stored in the buffer $\pqueue$ the 
process first erases it from the buffer in Line~\ref{ccanc:line4}.
Afterwards it checks whether the CAN driver needs to be informed as well. 
This check is performed by analysing $\txmirror$, which keeps the status of the TX buffers. 
In case there is a CAN message with identifier $\cid$ in TX buffer $\txid$, the actions of Lines~\ref{ccanc:line8}--\ref{ccanc:line13} are taken.

If the process {\CCANC} has sent a cancellation request before, i.e., the flag $\txab{\txid}{\txmirror}$ is true, 
the process does not take any action, and returns to the main process (Line~\ref{ccanc:line9}).
Otherwise the flag is set to true (Line~\ref{ccanc:line11}) and a cancellation message is sent to the CAN driver (Line~\ref{ccanc:line12}).

If the message with CAN identifier $\cid$ can be found in {\pqueue} but not in any of the TX buffers (in \txmirror), no 
cancellation request needs to be forwarded and the cancellation
process is finished (the message was not yet forwarded to the CAN
driver). As a consequence the process informs the instance of the fragmentation protocol responsible for this message
about the successful cancellation, by sending an \ackID-message in Line~\ref{ccanc:line16}.

\myparagraph{Notification from the CAN Driver} 
When an \ackID-message is received from the CAN driver by the multiplexer, the process \ACKC\ is executed.

  \algsetup{linenodelimiter=.,linenosize=\tiny}
  \begin{algorithm}[ht]
    {\footnotesize
      \caption{Acknowledgement from the CAN driver}
      \label{pro:ackc}
      \begin{algorithmic}[1]
        % !TEX root = ../paper.tex
{
\renewcommand{\newtask}[1]{\fnnewtask\ensuremath{(\nosp{#1},\nosp{\txmirror})}}
\DEFPROCESS{\ACKC}{\ackb\comma\txid\comma\pqueue\comma\txmirror}														\label{ackc:line0}
	\IF[mailbox idle]{$\txcid{\txid}{\txmirror} = \cidundef$}																						\label{ackc:line1}
		\errL																																							\label{ackc:line2}
	\ELSIF[mailbox non-idle, distill {\cid}]{$\txcid{\txid}{\txmirror} =\cid \wedge \cid \not=\cidundef$}						\label{ackc:line3}
		\UPD{\mboxupd{\txid}:=(\cidundef,\false)}																									\label{ackc:line4}
		\PAR																																							\label{ackc:line5}
		\IF	[positive acknowledgment received ]{$\ackb=\true\ \vee\ \pqueueupd{\cid} = \msgundef$} 						\label{ackc:line6}
			\UPD{\pqueueupd{\cid} := \msgundef}\COM{erase message }																\label{ackc:line7}
			\unicastL{\fragname{\cid}}{\ackM\ackb}\ .																								\label{ackc:line8}
			\PAR																																						\label{ackc:line9}
				\IF[more messages to be handled]{$\newtask{\pqueue} = \bid \wedge \bid\not=\cidundef$}					\label{ackc:line10}
					\UPD{\mboxupd{\txid}:=(\bid,\false)}																								\label{ackc:line11}
					\unicastL{\C}{\amsg{\txid}{\pqueueupd{\bid}}}\ .\											 									\label{ackc:line12}
					\multiL{\pqueue}{\txmirror}																												\label{ackc:line13}
				\ELSIF[nothing to handle at the moment]{$\newtask{\pqueue}=\cidundef$}											\label{ackc:line14}
					\multiL{\pqueue}{\txmirror}																												\label{ackc:line15}
				\ENDIFii
			\ENDPAR																																				\label{ackc:line16}
		\ELSIF[task needs rescheduled]{$\ackb=\false\ \wedge\ \pqueueupd{\cid}\not= \msgundef$}						\label{ackc:line17}
			\UPD{\bid := \newtask{\pqueue}}\COM{determine next task}																	\label{ackc:line18}
			\UPD{\mboxupd{\txid}:=(\bid,\false)}																										\label{ackc:line19}
			\unicastL{\C}{\amsg{\txid}{\pqueueupd{\bid}}}\ .\ 																					\label{ackc:line20}
			\multiL{\pqueue}{\txmirror}																														\label{ackc:line21}
		\ENDIFii
		\ENDPAR																																					\label{ackc:line22}
	\ENDIFii
}
	\end{algorithmic}
    }%end{footnotesize}
  \end{algorithm}

If the multiplexer received a message that carries an identifier of the TX buffer ($\txid$), but has no local knowledge about it  ($\txcid{\txid}{\txmirror}$ is undefined), 
something went wrong and the protocol deadlocks with an error. In the future we hope to show that this situation cannot occur. 

Lines~\ref{ackc:line3}--\ref{ackc:line22} present the standard case---an acknowledgement from \txid\ is received, informing about the status of a message with CAN ID $\cid$.
Independent of the contents of the acknowledgement, the TX buffer is erased (Line~\ref{ackc:line4}). 
By this the array $\txmirror$ is adapted to the status of the actual TX buffers of the CAN driver---the message with ID \cid\ was either sent successfully or erased from the buffer. 

If the sending of the CAN message was successful ($\ackb\mathop=\true$, Line~\ref{ackc:line6}),
the message is erased from the priority queue (Line~\ref{ackc:line7}), and
 the corresponding instance of the fragmentation
protocol (the one which is responsible for CAN ID \cid,
and determined by $\fragname{\cid}$) 
is informed about the success (Line~\ref{ackc:line8}).
Since the TX buffer $\txid$ is now empty, the process checks whether there are pending messages that needs to be scheduled. 
If $\newtask{\pqueue,\nosp{\txmirror}}=\cidundef$
(Line~\ref{ackc:line14}) no message needs to be scheduled and the 
process returns to the main process \MULT, where it can receive new, incoming messages. 
In case there is at least one message stored in $\pqueue$ that is not yet forwarded
to the CAN controller, the process chooses the message with best (least) CAN ID, using the function
$\newtask{\pqueue,\nosp{\txmirror}}$;
the least CAN ID is stored in the variable $\bid$. 
This CAN message is then unicast to the controller, together with the name $\txid$ of the TX buffer to be used (Line~\ref{ackc:line12});
the local knowledge about $\txid$ is adapted in
Line~\ref{ackc:line11}: the CAN ID $\bid$ is stored together with the
flag true, indicating that no cancel request are sent for this message.

In case of a negative acknowledgement ($\ackb\mathop=\false$), it matters whether the unsent CAN
message (with identifier $\cid$) still occurs in the priority queue ($\pqueueupd{\cid}\not= \msgundef$).
If this is the case (Line~\ref{ackc:line17}), the message must have been removed from the TX
buffer in order to make place for a message with a higher priority. Since a TX buffer is now
empty, a new CAN message $\bid$ can be passed to the CAN driver (if there is any). 
This scheduling happens in Lines~\ref{ackc:line18}--\ref{ackc:line21}, and is similar to Lines~\ref{ackc:line10}--\ref{ackc:line13}.

If the message $\cid$ that was dropped from the TX buffer does not occur in the priority queue
(Line~\ref{ackc:line6}), the message must have been removed from the TX buffer in response to a
cancellation request from the fragmentation protocol. The latter is informed about the successful
cancellation in Line~\ref{ackc:line8}. Subsequently, the multiplexer schedules another message for
transmission over the CAN bus if appropriate \mbox{(Lines~\ref{ackc:line9}--\ref{ackc:line16}).}

\myparagraph{Message Queue}
To guarantee that our system is non-blocking, the multiplexer is equipped with an in-queue. 
This input enabled queue (\Pro{queue}) runs in parallel with {\MULT}.
Every incoming message from the fragmentation protocol or the CAN driver
is first stored in this queue, and piped from
there to the multiplexer {\MULT}, whenever {\MULT} is ready to handle a
new message. 

Here we simply assume that the length of the queue is unbounded. However,
in the future we hope to show that under some mild conditions a queue with a limited capacity is
equally effective.

The process is identical to the CAN receiver (\Pro{can_receive} of \autoref{sec:can_spec}), except that the unicast-procedure is replaced by a {\bf send}-action, which triggers the forwarding to the process \MULT.

  \algsetup{linenodelimiter=.,linenosize=\tiny}
  \begin{algorithm}[ht]
    {\footnotesize
      \caption{Message Queue}
      \label{pro:queue}
      \begin{algorithmic}[1]
        % !TEX root = ../paper.tex
\DEFPROCESS{\QMSG}{\msgs}
	\IFempty
		\COMLINE{store incoming message at the end of \msgs}		
		\receiveL{\msg}\ . 			
		\qmsg{\append{\msg}{\msgs}}					
	\ELSIF[the queue is not empty]{$\msgs\not=[\,]$}			
		\PAR
		\COMLINE{pop top message and send it to the multiplexer}		
		\sendL{\head{\msgs}}\ .\ 
		\qmsg{\tail{\msgs}}												
		\COMLINE{or receive and store an incoming message}				
		\STATE $+$\,  \receive{\msg}\ . \qmsg{\append{\msg}{\msgs}} \label{qmsg:line8}        
		\ENDPAR
	\ENDIFii

	\end{algorithmic}
    }%end{footnotesize}
  \end{algorithm}

\noindent
The {\bf receive}-action in Line~\ref{qmsg:line8} is needed to guarantee 
input-enabledness, meaning that the process {\QMSG} is always ready to
receive a new, incoming message, even when the process is about to send a message itself.

\subsubsection{Initialisation}\label{ssec:multi_init}
The multiplexer $M_H$ is initialised by 
$\M\mathop{:}\big((\xi,\multi{\pqueue}{\txmirror})\parl
(\xii,\qmsg{\msgs})\big)$, 
where
$\xi(\pqueueupd{\cid}) = \msgundef$  for all \cid,
$\xi(\mboxupd{\txid}) = (\cidundef, \false)$ for all \txid, and 
$\xii(\msgs)=[\,]$.

\end{document}